\documentclass{ws-procs9x6}            

\usepackage{gt15,defs_lec16,gtRashba}
\begin{document}
\title{
Theory of electron transport and magnetization dynamics in metallic ferromagnets
}

\author{Gen Tatara}

\address{RIKEN Center for Emergent Matter Science (CEMS)\\
2-1 Hirosawa, Wako, Saitama 351-0198, Japan\\
$^*$E-mail: gen.tatara@riken.jp}

\begin{abstract}
Magnetic electric effects in ferromagnetic metals are discussed from the viewpoint of effective spin electromagnetic field that couples to conduction electron spin.
The effective field in the adiabatic limit is the spin Berry's phase in space and time, and it leads to spin motive force (voltage generated by magnetization dynamics) and topological Hall effect due to spin chirality.
Its gauge coupling to spin current describes the spin transfer effect, where magnetization structure is driven by an applied spin current.
The idea of effective gauge field can be extended to include spin relaxation and Rashba spin-orbit interaction.
Voltage generation by the inverse Edelstein effect in junctions is interpreted as due to the electric component of Rashba-induced spin gauge field.
The spin gauge field arising from the Rashba interaction turns out to coincides with troidal moment, and causes asymmetric light propagation (directional dichroism) as a result of the Doppler shift.
Rashba conductor without magnetization is shown to be natural metamaterial exhibiting negative refraction.
\end{abstract}


\bodymatter

\section{Introduction}

Our technology is based on various electromagnetic phenomena.
For designing electronics devices, the Maxwell's equation is therefore of essential importance.
The mathematical structure of the electromagnetic field is governed by a U(1) gauge symmetry, i.e.,  an invariance of physical laws under phase transformations.
The gauge symmetry is equivalent to the conservation of the electric charge, and was established when a symmetry breaking of unified force occured immediately after the big bang. 
The beautiful mathematical structure of charge electromagnetism was therefore determined when our universe started, and there is no way to modify its laws.

Interestingly, charge electromagnetism is not the only electromagnetism allowed in the nature. 
In fact, electromagnetism arises whenever there is a U(1) gauge symmetry associated with conservation of some effective charge.
In solids, there are several systems which have the U(1) gauge symmetry as a good approximation. 
Solids could thus display several types of effective electromagnetic fields.
A typical example is a ferromagnetic metal. 
In ferromagnetic metals, conduction electron spin (mostly $s$ electron) is coupled to the magnetization (or localized spins of $d$ electrons) by an interaction called the $sd$ interaction, which tends to align the electron spin parallel (or anti-parallel) to the localized spin. 
This interaction is strong in most 3$d$ ferromagnetic metals, and as a result, conduction electron's spin originally consisting of three components, reduces to a single component along the localized spin direction.
The remaining component is invariant under a phase transformation, i.e., 
has a U(1) gauge symmetry just like the electric charge does. 
A spin electromagnetic field thus emerges that couples to conduction electron's spin. 
%
\begin{figure}[tb]
\begin{center}
\includegraphics[width=0.5\hsize]{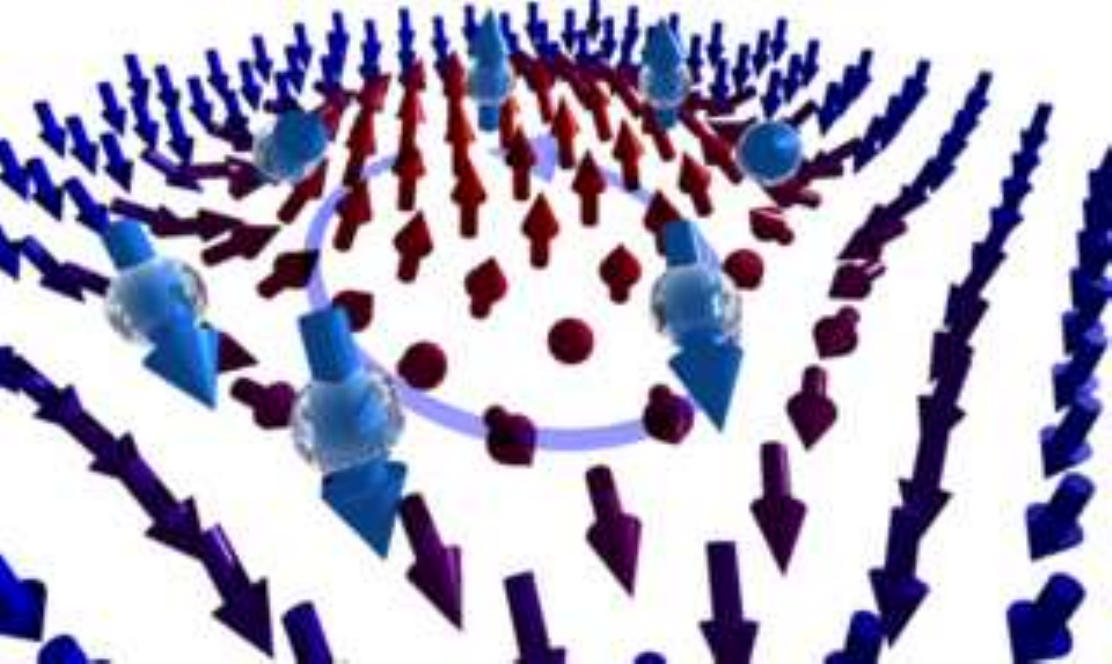}
\end{center}
\caption{
The spin of a conduction electron is rotated by a strong $sd$ interaction with magnetization as it moves in the presence of a magnetization texture, resulting in a spin gauge field. Magnetization texture is therefore equivalent to an effective electromagnetic field for conduction electron spin. 
}
\label{fig1}
\end{figure}

The subject of the present paper is this spin electromagnetic field.
Spin electromagnetic field drives electron's spin, and thus plays essential roles in spintronics.
There is a gauge field for the spin electromagnetic field, a spin gauge field, which couples to spin current of the conduction electron.
The gauge coupling describes the effects of spin current on the localized spin dynamics.
As we shall see, when a spin-polarized electric current is applied, the adiabatic spin gauge field leads to spin-transfer torque and moves the magnetization structure (Sec. \ref{SECmagdynamics}).
The world of spin electromagnetic field is richer than that of electric charge, since the electron's spin in solids is under influence of various interactions such as spin-orbit interaction.
We shall show that even magnetic monopoles can emerge (Sec. \ref{SEC:adiabatic})

A spin electromagnetic field was first discussed in the context of a voltage generated by a canting of a driven domain wall by L. Berger \cite{Berger86}, and mathematically rigorous formulation was given by G. Volovik \cite{Volovik87}.
The idea of effective gauge field was shown to be extended to the cases with spin relaxation \cite{Duine08}, and  Rashba interaction \cite{Takeuchi12,Kim12,Tatara_smf13,Nakabayashi14}. 

Some of the phenomena discussed in this paper overlaps those in the paper by R. Raimondi in this lecture series, studied base on the Boltzmann equation approach \cite{RaimondiLec17}.

\begin{figure}[tb]
  \begin{center}
    \includegraphics[width=0.4\hsize]{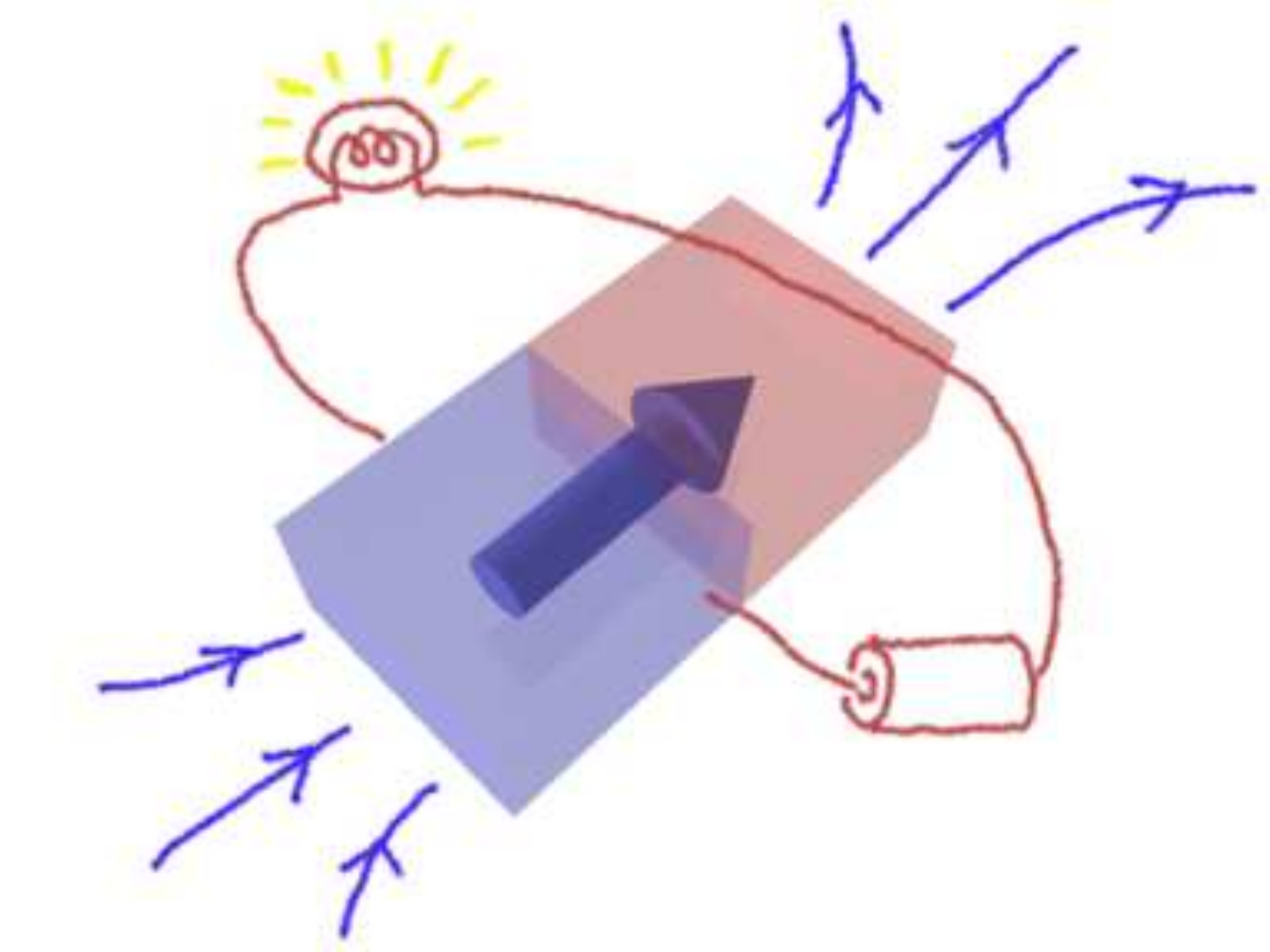}
  \end{center}
\caption{Ferromagnetic metals have  magnetization and conduct electricity,  indicating existence of  
localized spins and conduction electrons.
\label{FIGFM}}
\end{figure}

\section{Ferromagnetic metal}
Let us start with a brief introduction of ferromagnetic metals (Fig. \ref{FIGFM}).
Ferromagnets have magnetization, namely, an ensemble of localized spins.
Denoting the localized spin as $\Sv$, the magnetization is $\Mv=-\frac{\hbar \gamma}{a^3}\Sv$, where $\gamma(>0)$ and $a$ are gyromagnetic ratio and lattice constant, respectively.
As the electron has negative charge, the localized spin and magnetization points opposite direction.
In $3d$ transition metals, localized spins are aligned spins of $3d$ electrons.
Ferromagnetic metals have finite conductivity, indicating that there are conduction electrons, mainly $4s$ electrons.
The conduction electrons and $d$ electrons are coupled via $sd$ mixing.
As a result, there arises an exchange interaction between conduction electron spin, $\sev$,  and localized spin, which reads
\begin{align}
\Hsd=-\Jsd \Sv\cdot\sev,\label{Hsd}
\end{align}
where $\Jsd$ represents the strength.
In this article, the localized spin is treated as classical variable, neglecting the  conduction of $d$ electrons.

The dynamics of localized spin is described by the Landau-Lifshiz-Gilbert (LLG) equation,
\begin{align}
 \dot{\nv}&=\gamma\Bv\times\nv+\alpha\nv\times\dot{\nv}, \label{LLG}
\end{align}
where $\nv\equiv \Sv/S$ is a unit vector representing the direction of localized spin, 
$\Bv$ is the total magnetic field acting on the spin.
The last term of the right hand side represents the relaxation (damping) of localized spin, called the Gilbert damping effect and $\alpha$ is the Gilbert damping constant.
The Gilbert damping constant in most metallic ferromagnets are of the order of $10^{-2}$.

We shall now start studying phenomena arising from the exchange interaction, Eq. (\ref{Hsd}), between localized spin and conduction electron.

\begin{figure}[tbh]
\begin{center}
\includegraphics[width=0.3\hsize]{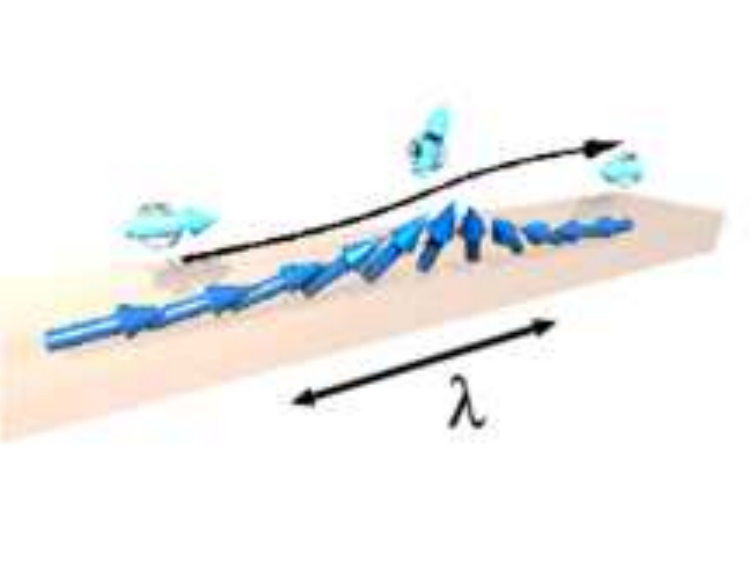}\hspace{8mm}
\includegraphics[width=0.3\hsize]{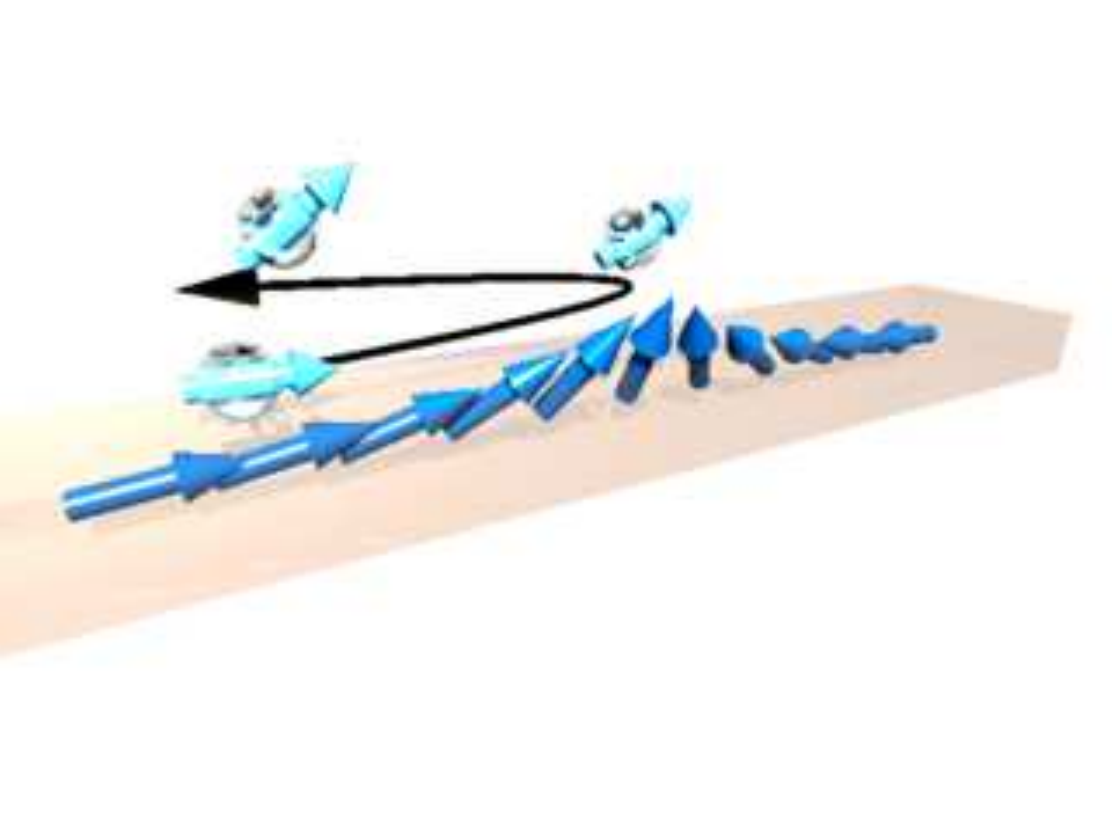}
\end{center}
\caption{Schematic figures showing conduction electron injected to a domain wall. 
(a): In the adiabatic limit, i.e., for a large domain wall width, the electron goes through the wall with a spin flip (left). 
(b): Non adiabaticity due to finite domain wall width leads to reflection and electric resistance (right).
\label{FIGDWelec}}
\end{figure}

\section{Electron transport through magnetic domain wall : phenomenology }

\begin{figure}[tb]
\begin{center}
\includegraphics[width=0.4\hsize]{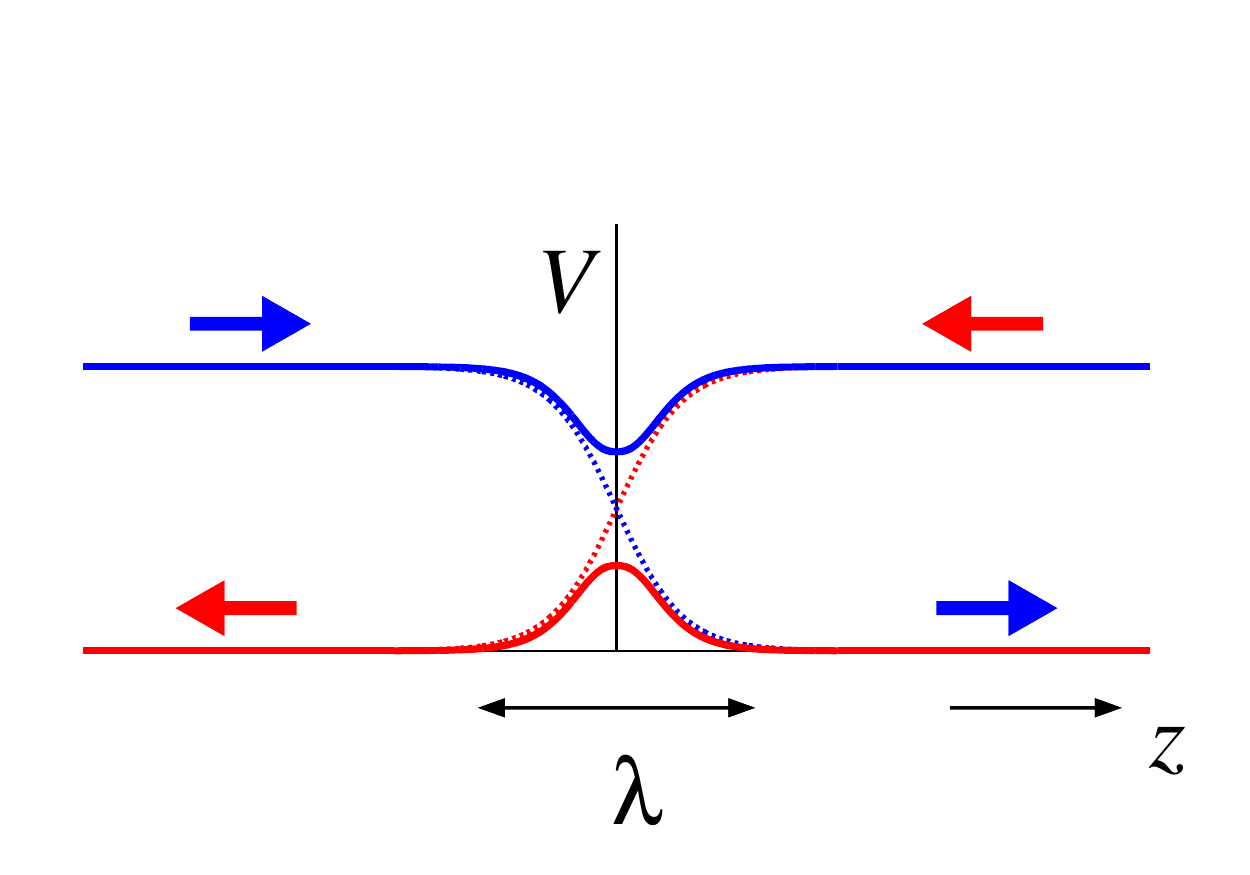}
\end{center}
\caption{Potential energy $V(z)$ for conduction electron with spin   $\ra$ and $\la$ as a result of $sd$ exchange interaction. Dotted lines are the cases neglecting spin flip inside the wall, while solid lines are with spin flip. 
\label{FIGDWpotentialenergy}}
\end{figure}

We consider a ferromagnetic domain wall, which is a structure where localized spins (or magnetization) rotate spatially (Fig. \ref{FIGDWelec}).
Its thickness, $\lambda$, in typical ferromagnets is $\lambda=10-100$nm.
Let us consider here what happens when a conduction electron goes through a domain wall.
The wall is a macroscopic object for electrons, since thickness is much larger than the typical length scale of electron, the Fermi wavelength, $1/\kf$, which is atomic scale in metals.
The electron is interacting with localized spin via the $sd$ exchange coupling, Eq. (\ref{Hsd}).
We consider the case of  positive $\Jsd$, but the sign does not change the scenario.
The $sd$ interaction tends to align parallel the localized spin and conduction electron spin. 
If localized spin is spatially uniform, therefore,  the conduction electron is also uniformly polarized, and electron transport and magnetism are somewhat decoupled.
Interesting effects arise if the localized spins are spatially varying like the case of a domain wall.
We choose the $z$ axis along the direction localized spins change.
The lowest energy direction (magnetic easy axis) for localized spins is chosen as along $z$ axis.
(The mutual direction between the localized spin and direction of spin change is irrelevant in the case without spin-orbit interaction.)  
The wall in this case is with localized spins inside the wall changing within the plane of localized spin, and such wall is called the  N\`eel wall. 
At $z=\infty$ the localized spin is $S_z=S$, and is $S_z=-S$ at $z=-\infty$, and those states are represented a $\ra$ and $\la$, respectively.
For $\leftarrow$ electron, the potential in the left regime is low because of $sd$ exchange interaction, while that in the right region is high (dotted lines in Fig. \ref{FIGDWpotentialenergy}).
That is, the localized spin structure due to a domain wall acts as a spatially varying magnetic field, resulting in  potential barriers, 
$V_\rightarrow(z)=-\Jsd S_z(z)$ and $V_\leftarrow(z)=\Jsd S_z(z)$.
Considering the domain wall centered at $x=0$ having profile of 
\begin{align}
 S_z(z)&=S\tanh\frac{z}{\lambda}, \;\;\;  S_x(z)=\frac{S}{\cosh\frac{z}{\lambda}},\;\;\;  S_y=0 ,
 \label{DWsolrest}
\end{align}
conduction electron's Schr\"odinger equation with energy $E$ reads 
\begin{align}
 \lt[-\frac{\hbar^2}{2m}\frac{d^2}{dz^2}-\Jsd S
  \lt(\sigma_z \tanh\frac{z}{\lambda}+\sigma_x \frac{1}{\cosh\frac{z}{\lambda}}\rt) \rt]\Psi=E\Psi,
 \label{elecSeqinDW}
\end{align}
$\Psi(z)=(\Psi_\rightarrow(z),\Psi_\leftarrow(z))$ begin the two-component wave function.
If the spin direction of the conduction electron is fixed along the $z$ axis, the potential barrier represetned by the term proportional to $\sigma_z$ leads to reflection of electron, but in reality, the electron spin can rotate inside the wall as a result of the term proportional to $\sigma_x$ in Eq. (\ref{elecSeqinDW}).
The mixing of $\leftarrow$ and $\rightarrow$ electron leads to the smooth potential barrier plotted as solid lines in Fig. \ref{FIGDWpotentialenergy}.

Let us consider an incident $\leftarrow$ electron from the left.
If the electron is slow, the electron spin can keep the lowest energy state by gradually rotating its direction inside the wall.
This is the adiabatic limit.
As there is no potential barrier for the electron in this limit, no reflection arises from the domain wall, resulting in a vanishing resistance  (Fig. \ref{FIGDWelec}(a))
In contrast, if the electron is fast, the electron spin cannot follow the rotation of the localized spin, resulting in a reflection and finite resistance (Fig. \ref{FIGDWelec}(b)).
The condition for slow and fast is determined by the relation between the time for the electron to pass the wall and the time for electron spin rotation. 
The former is $\lambda/\vf$ for electron with Fermi velocity $\vf(=\hbar\kf/m)$ 
(spin-dependence of the Fermi wave vector is neglected  and $m$ is the electron mass).
The latter time is $\hbar/\Jsd S$, as the electron spin is rotated by the $sd$ exchange interaction in the wall.
Therefore, if 
\begin{align}
\frac{\lambda}{\vf}\gg \frac{\hbar}{\Jsd S} , \label{adiabaticcondition}
\end{align}
is satisfied, the electron is in the adiabatic limit \cite{Waintal04}.
The condition of adiabatic limit here is the case of clean metal (long mean free path);
 In dirty metals, it is modified \cite{Stern92,TKS_PR08}.

The transmission of electron through a domain wall was calculated by G. G. Cabrera and L. M. Falicov \cite{Cabrera74}, and its physical aspects were discussed by L. Berger \cite{Berger78,Berger86}.
Linear response formulation and scattering approach were presented in Refs. \cite{TF97,GT00,GT01}.
The adiabaticity condition was discussed by  X. Waintal and M. Viret\cite{Waintal04}.

\subsection{Spin-transfer effect \label{SECstt}}

\begin{figure}[tb]
\begin{center}
\includegraphics[width=0.3\hsize]{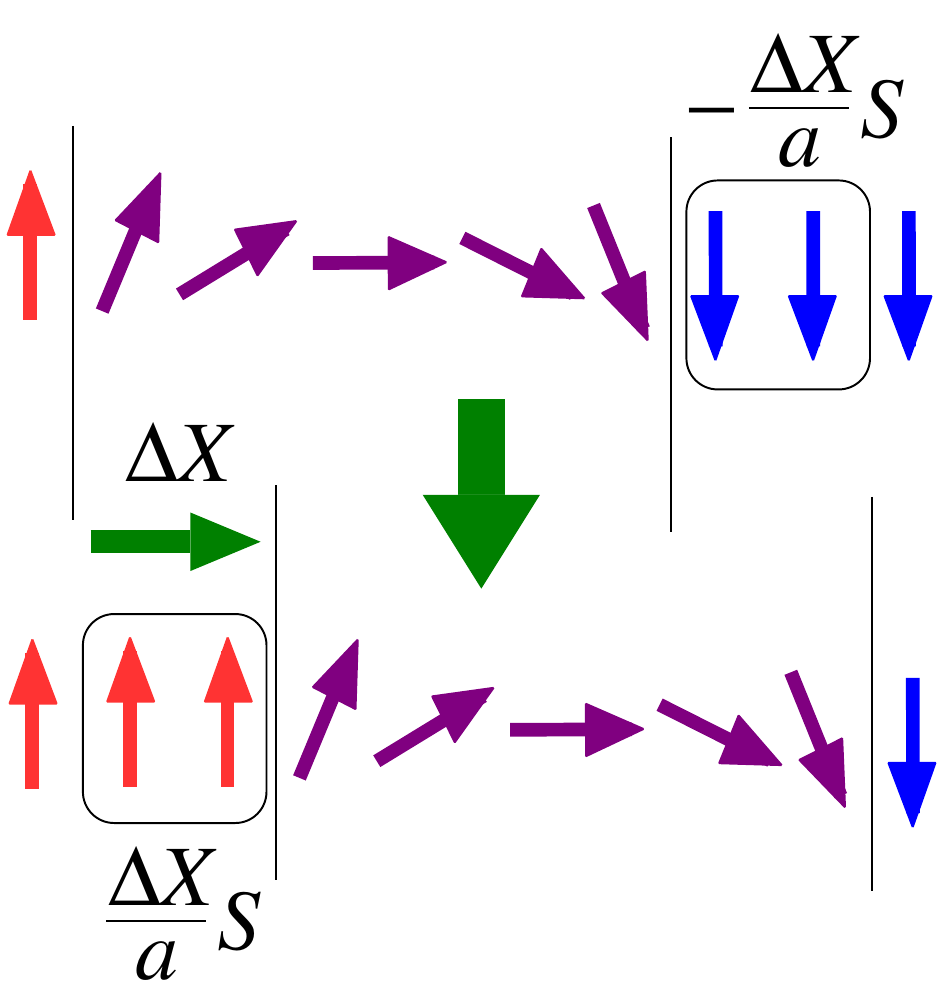}
\end{center}
\caption{ The shift of the domain wall by a distance $\Delta X$ results in a change of the spin of the localized spins $\frac{\Delta X}{a}S-\lt(-\frac{\Delta X}{a}S\rt)=2S\frac{\Delta X}{a}$.
The angular momentum change is therefore $\hbar$ if $\Delta X=\frac{a}{2S}$. 
\label{FIGdwdisplacement}}
\end{figure}

As we discussed above, in the adiabatic limit, the electron spin gets rotated after passing through the wall (Fig. \ref{FIGDWelec}(a)).
The change of spin angular momentum, $2\times\frac{\hbar}{2}=\hbar$, must be absorbed by the localized spins.
(Angular momentum dissipation as a result of spin relaxation is slow compared to the exchange of the angular momentum via the $sd$ exchange interaction.)
To absorb the spin change of $\hbar$, the domain wall must shift to the right, resulting in an increase of the spins $\leftarrow$.
We consider for simplicity the case of cubic lattice with lattice constant $a$.
The distance of the wall shift $\Delta X$  necessary to absorb the electron's spin angular momentum of $\hbar$ is then $[\hbar/(2\hbar S)]a$ (Fig. \ref{FIGdwdisplacement})).
If we apply a spin-polarized current through the wall with the density $j_{\rm s}$
(spin current density is defined to have the same unit of A/m$^2$ as the electric current density.)
The rate of the angular momentum change of the conduction electron per unit time and area is 
$\hbar j_{\rm s}/e$.
As the number of the localized spins in the unit area is $1/a^2$, the wall must keep moving a distance of 
$(j_{\rm s}/e)(a^3/2S)$ per unit time.
Namely, when a spin current density is applied, the wall moves with the speed of 
\begin{align}
v_{\rm s}\equiv \frac{a^3}{2eS} j_{\rm s} . \label{vst}
\end{align}
This effect was pointed out by L. Berger \cite{Berger86} in 1986, and is now called the spin-transfer effect after the papers by J. Slonczewski \cite{Slonczewski96}.

From the above considerations  in the adiabatic limit, we have found that a domain wall is driven by  spin-polarized current, while the electrons do not get reflected and no resistance arises from the wall.
These two facts naively seem inconsistent, but are direct consequence of the fact that  a domain wall is a composite structure having both linear momentum and angular momentum.
The adiabatic limit is the limit where angular momentum is transfered between the electron and the wall, while no linear momentum is transfered.

\begin{figure}[tb]
\begin{center}
\includegraphics[width=0.27\textwidth]{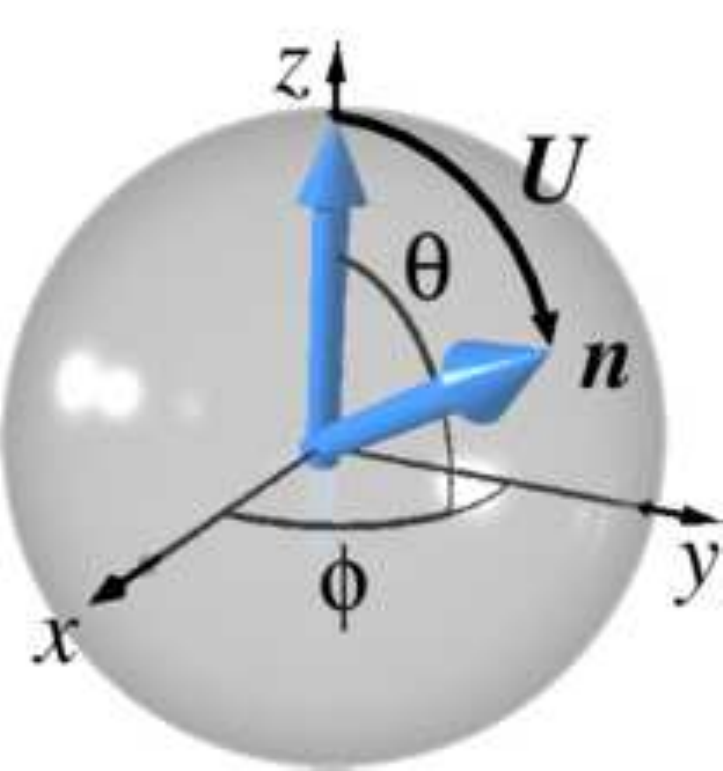}\hspace{12mm}
\includegraphics[width=0.27\textwidth]{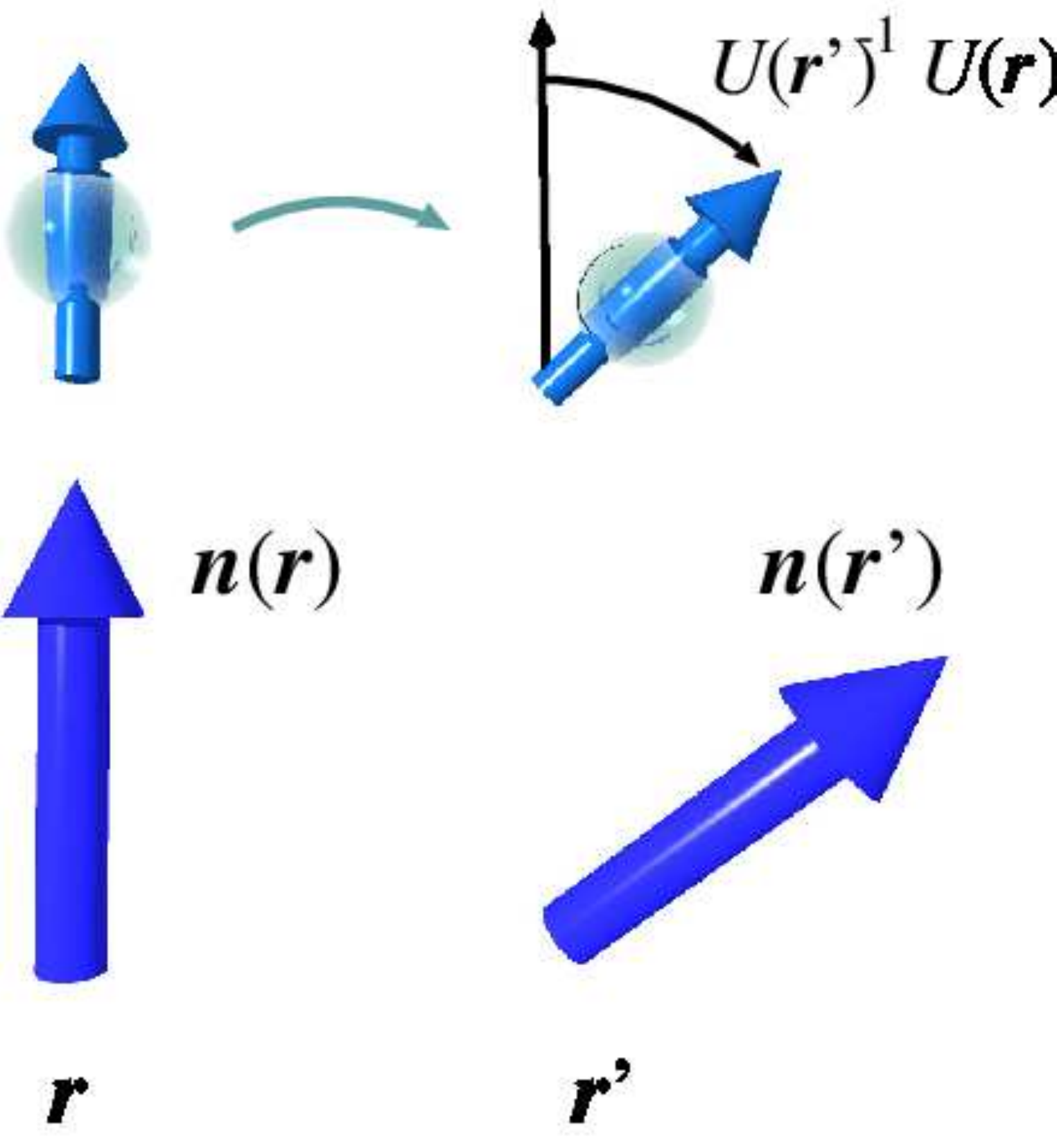}
 \caption{
Left: A Unitary transformation $U(\theta,\phi)$ relates the two spin configurations $|\!\uparrow\rangle$ and 
$|\nv\rangle$ as $|\nv\rangle=U|\!\uparrow\rangle$.
Right: The overlap of the wave functions at sites $\rv$ and $\rv'$ is 
$\langle \nv(\rv)|\nv(\rv')\rangle=\langle \uparrow\!|U(\rv')^{-1} U(\rv)|\!\uparrow\rangle$.
\label{FIGU}}
\end{center}
\end{figure}
\section{Adiabatic phase of electron spin \label{SEC:adiabatic}}
Transport of conduction electrons in the adiabatic (strong $sd$) limit is theoretically studied by calculating the quantum mechanical phase attached to the wave function of electron spin. 
We here consider a conduction electron hopping from a site $\rv$ to a neighboring site at $\rv'\equiv \rv+\av$ 
($\av$ is a vector connecting neighboring sites)(Fig. \ref{FIGU}).
The localized spin direction at those sites are $\nv(\rv)\equiv\nv$ and $\nv(\rv+\av)\equiv\nv'$, respectively, and the electron's wave function at the two sites are  
\begin{align}
|\nv\rangle &=\cos\frac{\theta}{2}|\!\uparrow\rangle+\sin\frac{\theta}{2}e^{i\phi}|\!\downarrow\rangle \nnr 
|\nv'\rangle &=\cos\frac{\theta'}{2}|\!\uparrow\rangle+\sin\frac{\theta'}{2}e^{i\phi'}|\!\downarrow\rangle,
\end{align}
where $\theta$, $\phi$ and $\theta'$, $\phi'$  are the polar angle of $\nv(\rv)$ and $\nv(\rv')$, respectively (Fig. \ref{FIGU}).
The wave functions are concisely written by use of matrices, $U(\rv)$ and $U(\rv')$, which rotates the spin state $|\!\uparrow\rangle$ to $|\nv\rangle$ (Fig. \ref{FIGU}), as  $|\nv\rangle = U(\rv)|\!\uparrow\rangle$ and $|\nv'\rangle = U(\rv')|\!\uparrow\rangle$. 
The rotation matrix is given by \cite{Sakurai94} (neglecting irrelevant phase factors)
\begin{align}
   U(\rv)=e^{\frac{i}{2}(\phi-\pi)\sigma_z}e^{\frac{i}{2}\theta\sigma_y}e^{-\frac{i}{2}(\phi-\pi)\sigma_z}
= \lt(\begin{array}{cc} \cos\frac{\theta}{2} & \sin\frac{\theta}{2}e^{i\phi} \\
        -\sin\frac{\theta}{2}e^{-i\phi} & \cos\frac{\theta}{2} \end{array} \rt) .
\label{Udef}
\end{align}
The overlap of the electron wave functions at the two sites is thus 
$\langle \nv'|\nv\rangle= \langle \uparrow\!|U(\rv')^{-1} U(\rv)|\!\uparrow\rangle$.
When localized spin texture is slowly varying, we can expand
the matrix product with respect to $\av$ as  $U(\rv')^{-1}U(\rv)=1-U(\rv)^{-1}(\av\cdot\nabla)U(\rv)+O(a^2)$ to obtain 
\begin{align}
\langle \nv'|\nv\rangle \simeq 1-\langle \uparrow\!|U(\rv)^{-1}(\av\cdot\nabla)U(\rv)|\!\uparrow\rangle
\simeq e^{i\varphi},
\label{phasedef}
\end{align}
where 
\begin{align}
\varphi\equiv i\av\cdot\langle\uparrow|U(\rv)^{-1}\nabla U(\rv)|\uparrow\rangle\equiv \av\cdot\Asv.
\end{align}
Since $(U^{-1}\nabla U)^\dagger =-U^{-1}\nabla U$, $\varphi$ is real.
A vector $\Asv$ here plays a role of a gauge field, similarly to that of the electromagnetism, 
and it is  called (adiabatic) spin gauge field. 
By use of Eq. (\ref{Udef}), this gauge field reads (the factor of $\frac{1}{2}$ represents the magnitude of electron spin)
\begin{align}
\Asv=\frac{\hbar}{2e}(1-\cos\theta)\nabla\phi.
\label{Asdef}
\end{align}

For a general path $C$, the phase is written as an integral  along $C$ as 
\begin{align}
\varphi=\frac{e}{\hbar} \int_C d\rv\cdot \Asv.
\end{align}
Existence of path-dependent phase means that there is an effective magnetic field, $\Bsv$, as seen by rewriting the integral over a closed path by use of the Stokes theorem as
\begin{align}
\varphi=\frac{e}{\hbar} \int_S d\Sv\cdot\Bsv,
\end{align}
 where
\begin{align}
\Bsv\equiv \nabla\times\Asv,
\end{align}
represents the curvature or effective magnetic field.
This phase $\varphi$, arising from strong $sd$ interaction, couples to electron spin, and is called the spin Berry's phase.  
Time-derivative of phase is equivalent to a voltage, and thus we have effective electric field defined by
\begin{align}
\dot{\varphi}=-\frac{e}{\hbar} \int_Cd\rv\cdot\Esv,
\end{align}
 where 
 \begin{align}
\Esv\equiv-\dot\Asv,
\end{align}
(For a gauge invariant expression of $\Esv$, we need to include the time component of the gauge field, $A_{{\rm s},0}$ \cite{Tatara12}.)
In terms of vector $\nv$ the effective fields read 
\begin{align}
{\Ev}_{{\rm s},i}&
             = -\frac{\hbar}{2e} \nv \cdot (\dot{\nv} \times \nabla_i \nv)     
                         \nnr
{\Bv}_{{\rm s},i}&= \frac{\hbar}{4e}{\sum}_{jk}\epsilon_{ijk} \nv \cdot (\nabla_j \nv \times \nabla_k \nv).
\label{EsBsdef}
\end{align}
\begin{figure}[bt]
\begin{center}
\includegraphics[width=0.3\textwidth]{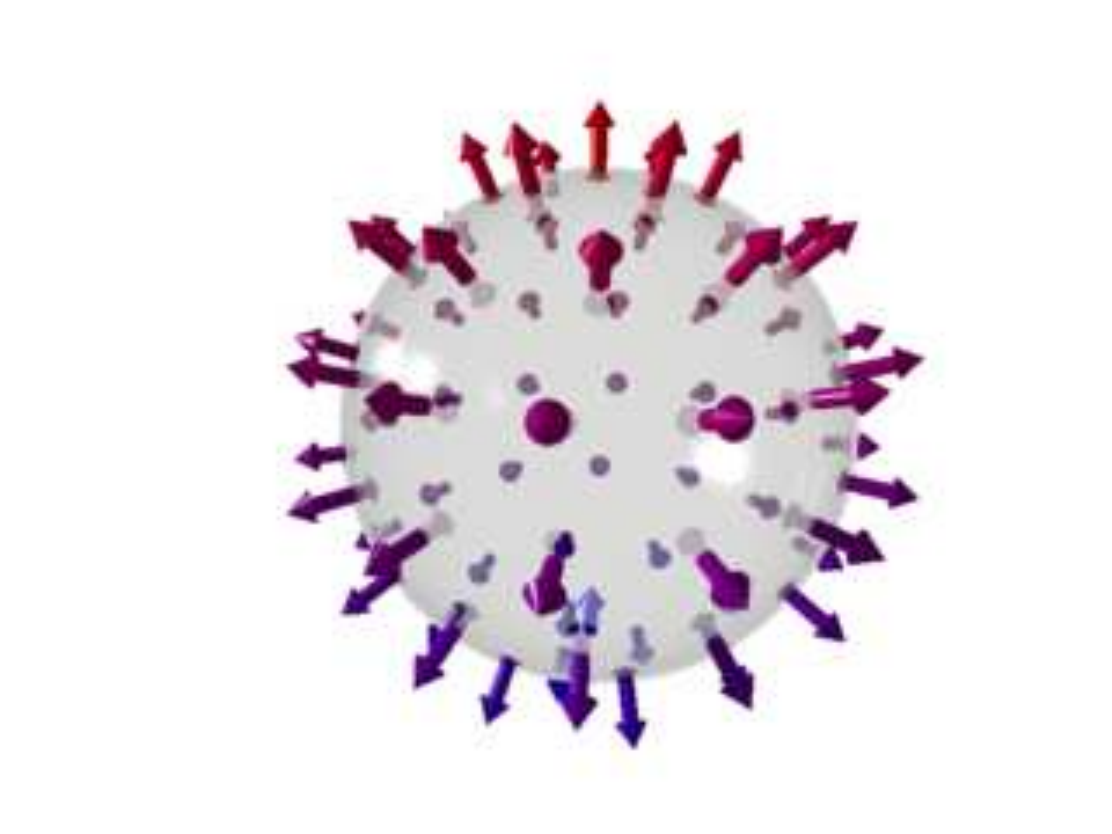}
\includegraphics[width=0.3\textwidth]{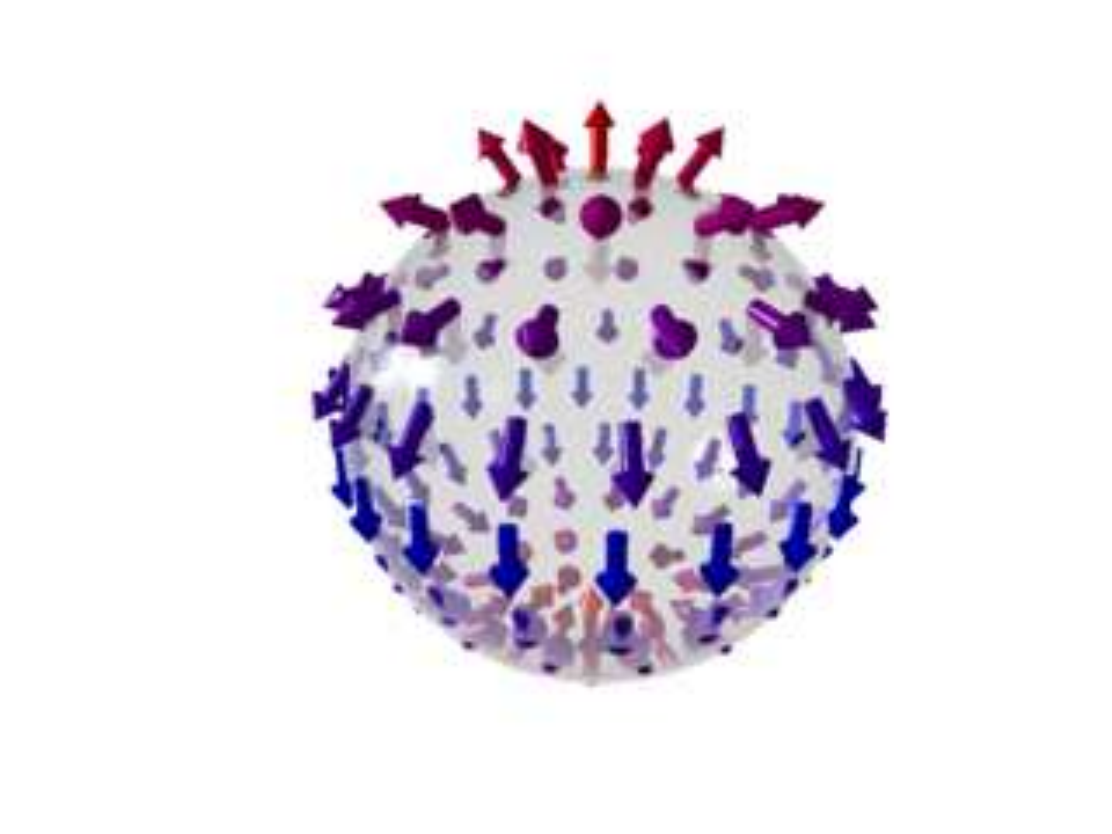}
\end{center}
\caption{
Magnetization structures, $\nv(\rv)$, of a hedgehog monopole having a monopole charge of $Q_{\rm m}=1$ and the one with $Q_{\rm m}=2$ .
At the center, $\nv(\rv)$ has a singularity and this gives rise to a finite monopole charge. 
}
\label{FIGHH}
\end{figure}
These two fields couple to the electron spin and are called spin electromagnetic fields ($\Asv$ is spin gauge field). They satisfy the Faraday's law,
\begin{align}
\nabla\times\Esv+\dot{\Bsv}=0,
\end{align}
  as a trivial result of their definitions. 
  Defining the spin magnetic charge as 
\begin{align}
\nabla\cdot\Bsv\equiv \rho_{\rm m},
\end{align}
we see that $\rho_{\rm m}=0$ as a local identity, since spin vector with fixed length has only two independent variables, and therefore 
${\sum}_{ijk}\epsilon_{ijk} (\nabla_i\nv) \cdot (\nabla_j \nv \times \nabla_k \nv)=0$.
However, there is a possibility that the volume integral, $Q_{\rm m}\equiv \intr \rho_{\rm m}$, is finite; In fact, using the Gauss's law we can write ($\int d\Sv $ represents a surface integral)
\begin{align}
Q_{\rm m} =\frac{h}{4\pi e}\int d\Sv\cdot {\bm \Omega},
\end{align}
 and  it follows that $Q_{\rm m} =\frac{h}{e}\times$integer since 
$\frac{1}{4\pi}\int d\Sv\cdot{\bm \Omega}$ is a winding number, an integer, of a mapping from a sphere in the coordinate space to a sphere in spin space. 
If the mapping is topologically non-trivial as a result of a singularity, the monopole charge is finite. 
Typical nontrivial structures of $\nv$ are shown in Fig.  \ref{FIGHH}.
The  singular structure with a single monopole charge is called the hedgehog monopole. 

The Faraday's law similarly reads
$ (\nabla\times\Esv)_i+\dot{\Bsv}_i=\frac{\hbar}{4e}\sum_{ijk}\epsilon_{ijk} \dot\nv \cdot (\nabla_j \nv \times \nabla_k \nv) \equiv \jv_{\rm m} $, which vanishes locally but is finite when integrated, indicating that topological monopole current $\jv_{\rm m}$ exists.

The other two Maxwell's equations describing $\nabla\cdot\Esv$ and $\nabla\times\Bsv$ are derived by evaluating the induced spin density and spin current based on linear response theory \cite{Takeuchi12,Tatara12}.

\begin{figure}[tb]
  \begin{center}
    \includegraphics[width=0.35\hsize]{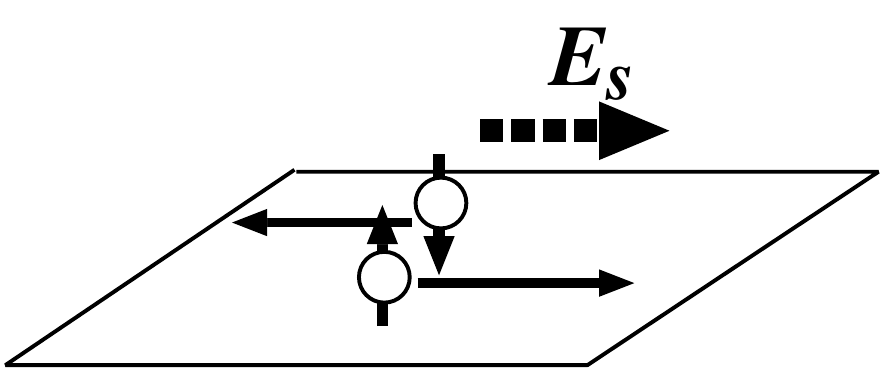}
    \includegraphics[width=0.35\hsize]{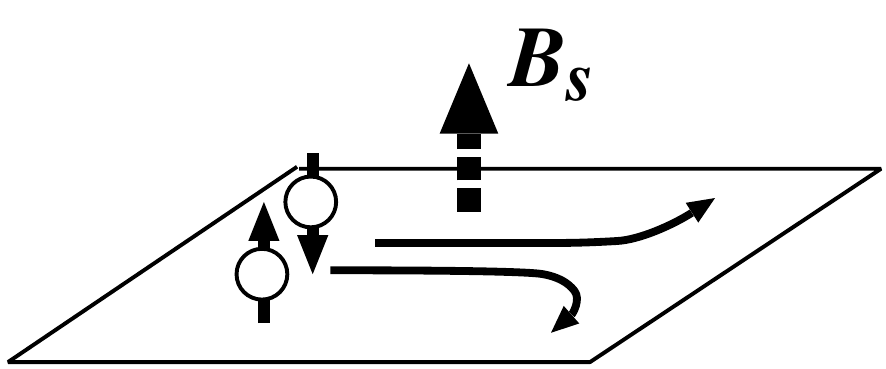}
  \end{center}
\caption{Spin electric field $\bm{E}_{\rm s}$ and spin magnetic field $\bm{B}_{\rm s}$ act oppositely for electrons with opposite spin, and thus are useful for generation of spin current.   
\label{FIGEsBs}}
\end{figure}

\section{Detection of spin electromagnetic fields}
The spin electromagnetic fields are real fields detectable in transport measurements.
They couples to the spin polarization of the electrons (Fig. \ref{FIGEsBs}), and because spin density and spin current in ferromagnetic metals is always accompanied with electric charge and current, respectively, 
 the effects of the spin magnetic fields are observable in electric measurements.
The electric component $\Esv$ is directly observable as a voltage generation from magnetization dynamics, and the voltage signals of $\mu$V order have been observed for the motion of domain walls and vortices \cite{Yang09,Tanabe12}.
The spin magnetic field causes an anomalous Hall effect of spin, i.e., the spin Hall effect called the topological Hall effect.
The spin electric field arises if magnetization structure carrying spin magnetic field becomes dynamical due to the Lorentz force from $\Bsv$ according to $\Esv=\vv\times\Bsv$, where $\vv$ denotes the electron spin's velocity.
The topological Hall effect due to skyrmion lattice turned out to induce Hall resistivity of 4n$\Omega$cm \cite{Neubauer09,Schulz12}.
Although those signals are not large, existence of spin electromagnetic fields is thus confirmed experimentally.
It was recently shown theoretically that spin magnetic field couples to helicity of circularly polarized light (topological inverse Faraday effect) \cite{Taguchi12}, and an optical detection is thus possible.

\section{Effects of spin gauge field on magnetization dynamics\label{SECmagdynamics}}
As discussed in the previous section, the spin gauge field are measured by transport experiments.
Here we study the opposite effect, the effects of spin gauge field on magnetization dynamics when spin current is applied.
The spin gauge field is expected to couple to the spin current of the electron, $\js$, via the minimal coupling,
\begin{equation}
  H_{A_{\rm s}}=  \intr\lt[ -\frac{\hbar}{e}\Asv\cdot \jsv + \frac{n\hbar^2}{2m} (\Asv)^2 
  -{2\hbar}  A_{{\rm s},0}\rhos \rt]  ,\label{HAs}
\end{equation}
where $n$ is the electron density, and $\rhos=\frac{1}{2}(n_+-n_-)$ is the electron spin density, $n_\sigma$ ($\sigma=\pm$) representing the density of electron with spin $\sigma$.
The field $ A_{{\rm s},0}$ is the time component of spin gauge field (Eq. (\ref{Asdef}) with spatial derivative replaced by time derivative).
 (For rigorous derivation of the coupling, see Eqs. (\ref{Lezexpression2})(\ref{gaugeH}).)
As the spin gauge field is written in terms of localized spin variables, $\theta$ and $\phi$, as a result of  Eq. (\ref{Asdef}), this interaction describes how the spin current and electron density affects the magnetization dynamics.
Here we study the adiabatic limit, where the contribution second order in $\Asv$ (the second term of the right hand side of Eq. (\ref{HAs}) is neglected. 
Including the gauge interaction, the Lagrangian for the localized spin reads 
\begin{align}
 L_S &= \intr \lt[ \frac{2}{a^3} A_{{\rm s},0}\lt(S+\rhos a^3\rt) - \frac{\hbar}{e} \Asv\cdot \jsv\rt] -H_S,
\label{LS1}
 \end{align}
where $H_S$ is the Hamiltonian.
We see that the magnitude of localized spin is modified to be the effective one $\Sbar\equiv S+{\rhos}a^3$ including the spin polarization of the conduction electron.
Writing the gauge field terms explicitly, we have 
\begin{align}
 L_S &= \intr \lt[ \Sbar (1-\cos\theta) \lt( \frac{\partial}{\partial t}-\vv_{\rm s}\cdot\nabla \rt) \phi\rt] -H_S, 
 \label{Lsjstot}
\end{align}
where $\vv_{\rm s}\equiv \frac{a^3}{2e\Sbar}\jsv$.
The velocity $\vv_{\rm s}$ here agrees with the phenomenological one, Eq. (\ref{vst}), if electron spin polarization is neglected (i.e., if $\Sbar=S$).
In the adiabatic limit, therefore, the time-derivative of the localized spin in the equation of motion is replaced by the Galilean invariant form with a moving velocity of $\vv_{\rm s}$ when a spin current is present.
The equation of motion derived from the Lagrangian (\ref{Lsjstot}) reads  
\begin{align}
  \lt( \frac{\partial}{\partial t} -\vv_{\rm s}\cdot\nabla \rt) \Sv 
   & = -\gamma \Bv_{S}\times\Sv ,   \label{LLmoving}
\end{align}
where $\Bv_{S}$ is the effective magnetic field due to $H_S$. 
From Eq. (\ref{LLmoving}), it is obvious that the magnetization structure flows with velocity $\vv_{\rm s}$, and this effect is in fact the spin-transfer effect discussed phenomenologically in Sec. \ref{SECstt}.
It should be noted that  the effect is mathematically  represented by a simple gauge interaction of Eq. (\ref{LS1}). 
The equation of motion (\ref{LLmoving}) is the Landau-Lifshiz-Gilbert (LLG) equation including adiabatic spin-transfer effect. 
It was theoretically demonstrated that the spin-transfer torque induces a red shift of spin wave, resulting in instability of uniform ferromagnetic state under spin-polarized current \cite{STK05}.

In reality, there is nonadiabatic contribution described by spin-flip interactions. Such contribution leads to a mixing of the electron spins  resulting in a scattering of the conduction electron and a finite resistance due to the magnetization structure \cite{GT00,GT01}. 
This scattering gives rise to a force on the magnetization structure as a counter action \cite{TK04}.

As we have seen, the concept of adiabatic spin gauge field is useful to give a unified description of both  
electron transport properties in the presence of magnetization structure and the magnetization dynamics in the presence of spin-polarized current.

\section{Field-theoretic description}
So far we discussed that an effective spin gauge field emerges by looking into the quantum mechanical phase factor attached to conduction electron in the presence of magnetization structures.
Existence of effective gauge field is straightforwardly  seen in field-theoretic description.

A field-theoretical description is based on the Lagrangian of the system, 
\begin{align}
\Lhat= i\hbar\intr\sum_{\sigma} \chat^\dagger_\sigma \dot{\chat}_\sigma  -\Hhat,      \label{Lhat}                                                                    \end{align}
where $\Hhat=\hat{K}+\Hhat_{sd}$ is the field Hamiltonian.
Here   
\begin{equation}
  \hat{K}= \intr \sum_\sigma \chat^\dagger_\sigma\lt(- \frac{\hbar^2}{2m} \nabla^2 \rt) \chat_\sigma 
    = \frac{\hbar^2}{2m} \sum_\sigma \intr (\nabla \chat^\dagger_\sigma) (\nabla \chat_\sigma)
\end{equation}
describes the free electron part in terms of field operators for conduction electron, $\chat_{\sigma}$ and $\chat^\dagger_{\sigma}$, where $\sigma=\pm$ denotes spin.
The $sd$ exchange interaction is represented by 
\begin{equation}
 \Hhat_{sd}=  -\frac{J_{sd}S}{2} \intr \chat^\dagger(\nv\cdot \sigmav) \chat.
\end{equation}
We are interested in the case where 
$\nv(\rv,t)$ changes in space and time slowly compared to the electron's momentum and energy scales.
How the electron 'feels' when flowing through such slowly varying structure is described by introducing a rotating frame where the $sd$  exchange interaction is locally diagonalized.
In Sec. \ref{SEC:adiabatic}, we introduced a unitary matrix $U(\rv,t)$, and this matrix is used here to introduce a new  electron operator as 
\begin{align}
\ahat(\rv,t)=U(\rv,t)\chat(\rv,t). \label{Ucdef}
\end{align}
The new operator $\ahat$ describes the low energy dynamics for the case of strong $sd$ exchange interaction. In fact,  the $sd$ exchange interaction for this electron is diagonalized to be 
\begin{equation}
 \Hhat_{sd}=  -M \intr \ahat^\dagger \sigma_z \ahat ,
  \label{Hsddiag}
\end{equation}
where $M\equiv \frac{J_{sd}S}{2}$.
Instead, the kinetic term for the new electron is modified, because derivative of the electron field is modified as 
\begin{align}
\nabla \chat 
=U(\nabla+i \Acals ) \ahat, \label{covariantderivative}
\end{align}
where 
\begin{align}
\Acals\equiv -iU^\dagger\nabla U .
\end{align}
Here $\Acals$ is a $2\times2$ matrix, whose componets are represented by using Pauli matrices as 
\begin{align}
\Acal_{{\rm s},i}=\sum_{\alpha=x,y,z}\Acal_{{\rm s},i}^\alpha \sigma_\alpha.
\end{align}
Equation (\ref{covariantderivative}) indicates that the new electron field $\ahat$ is interacting with an effective gauge field, $\Acals$.
This gauge field has three components, is non-commutative and is called the SU(2) gauge field.
The three components explicitly read
\begin{equation}
\vecth{\Acal_{{\rm s},\mu}^x}{\Acal_{{\rm s},\mu}^y}{\Acal_{{\rm s},\mu}^z} 
= \hf 
\vecth{
-\partial_\mu \theta \sin \phi -\sin\theta \cos\phi \partial_\mu \phi }{
\partial_\mu \theta \cos \phi -\sin\theta \sin\phi \partial_\mu \phi }{
  (1-\cos\theta)\partial_\mu \phi  }.
\label{Aexpression}
\end{equation}
Due to Eq. (\ref{covariantderivative}), the kinetic term $\hat{K}$ is  written in terms of $\ahat$ electron as
\begin{equation}
  \hat{K} = \frac{\hbar^2}{2m} \intr [(\nabla-i\Acals) \ahat^\dagger] [(\nabla+ i\Acals) \ahat].
\end{equation}
Similarly, time-component of the gauge field
\begin{align}
\Acal_{{\rm s},0}\equiv -iU^\dagger\partial_t U,
\end{align}
arises from the time-derivative term ($i\hbar \chat^\dagger_\sigma \dot{\chat}_\sigma  $) of the Lagrangian (\ref{Lhat}). 
The Lagrangian in terms of $\ahat$ electron therefore reads 
\begin{align}
 \Lhat &= \intr \left[
i\hbar \ahat^\dagger \dot{\ahat} -\frac{\hbar^2}{2m}|\nabla \ahat|^2 +\eF \ahat^\dagger \ahat
 +\spol \ahat^\dagger \sigma_z \ahat
  \right.\nonumber\\
 & \left.+i\frac{\hbar^2}{2m}\sum_{i}
 (\ahat^\dagger {\cal A}_{{\rm s},i} \nabla_i \ahat - (\nabla_i\ahat^\dagger) {\cal A}_{{\rm s},i} \ahat )
  -\frac{\hbar^2}{2m}{\cal A}_{\rm s}^2 \ahat^\dagger \ahat -\hbar \ahat^\dagger {\cal A}_{{\rm s},0} \ahat
 \right].
\label{Lezexpression1}
\end{align}
If we introduce electron density operator, $\hat{n}\equiv   \ahat^\dagger \ahat$, and operators for spin dnesity and spin current density as 
\begin{align}
 \hat{\rhos}_\alpha & \equiv \frac{1}{2} \ahat^\dagger \sigma_\alpha \ahat , &
 \hat{j}_{{\rm s},i}^{\alpha} \equiv \frac{-i}{2m} \ahat^\dagger \vvec{\nabla}_i \sigma_\alpha \ahat 
  \equiv \frac{-i}{2m}\lt[ \ahat^\dagger \sigma_\alpha ({\nabla}_i \ahat)- ({\nabla}_i \ahat^\dagger) \sigma_\alpha \ahat \rt] ,
\end{align}
it reads 
\begin{align}
 \Lhat &= \intr\left[
i\hbar \ahat^\dagger \dot{\ahat} -\frac{\hbar^2}{2m}|\nabla \ahat|^2 +\eF \ahat^\dagger \ahat
  + \spol \ahat^\dagger \sigma_z \ahat
  - \hat{j}_{{\rm s},i}^{\alpha} {\cal A}_{{\rm s},i}^\alpha
  -\frac{\hbar^2}{2m}{\cal A}_{\rm s}^2 \hat{n} 
   -   \hat{\rhos}^{\alpha} {\cal A}_{{\rm s},0}^\alpha
 \right].
\label{Lezexpression2}
\end{align}

In the case of $\spol/\ef\gg 1$ (large $\Jsd$), the electron with spin $\downarrow$ has high energy because of strong spin splitting, Eq. (\ref{Hsddiag}), and is neglected. In this case, only the $z$ component of the gauge field, $\Acal_{{\rm s},i}^z$, survives. This component is thus essentially a U(1) gauge field, which coincides with the U(1) gauge field we have obtained from the argument of electron's phase factor, namely, $\Asv={\bm \Acal}_{{\rm s}}^z$.
The total Hamiltonian in the limit of large $J_{sd}$ therefore reduces to the one for  a charged particle in the presence of  a U(1) gauge field $\Asv$;
\begin{equation}
  \Hhat = \intr \lt[ \frac{\hbar^2}{2m} [(\nabla-i\Asv) \ahat_\uparrow^\dagger] [(\nabla+ i\Asv) \ahat_\uparrow]
     -\frac{J_{sd}S}{2} \ahat_\uparrow^\dagger \ahat_\uparrow \rt].
     \label{gaugeH}
\end{equation}
The field-theoretic method present here is highly useful, as it leads to a conclusion of the existence of an effective gauge field for spin simply by carrying out a unitary transformation to diagonalize strong $sd$ exchange interaction.

\section{Non-adiabaticity and spin relaxation }

\begin{figure}[tb]
  \begin{center}
    \includegraphics[height=6\baselineskip]{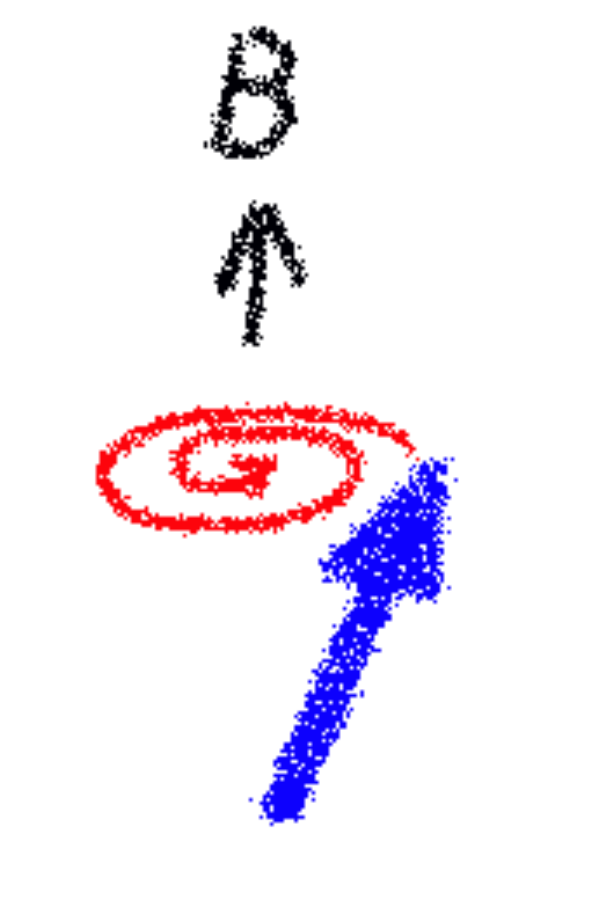}
  \end{center}
\caption{
Spin relaxation induces a torque perpendicular to the spin motion and let the spin relax to the stable direction along the external magnetic field.
\label{FIGnonadiabatic}}
\end{figure}

In reality, there is a deviation from the adiabatic limit we have considered so far.
One origin is the fact that the magnetization structure is not in the slowly varying limit, but has a finite length scale of spatial modulation. 
This effect, we call the non adiabaticity, leads to reflection of conduction electron by magnetization structures as in Fig. \ref{FIGDWelec}(b), resulting in a force on the magnetization structure when an electric current is applied \cite{Berger78,TK04}. 
In terms of torque, the effect of the force due to reflection is represented by a non-local torque, as it arises from finite momentum transfer \cite{TKSLL07}.
Another effect we need to take into account is the relaxation (damping) of spin schematically shown in the Fig. \ref{FIGnonadiabatic}. 
In metallic ferromagnets, the damping mostly arises from the spin-orbit interaction, as seen from the fact that the Gilbert damping parameter $\alpha$ and the $g$ value has a correlation of $\alpha\propto (g-2)^2$  as shown in Ref. \cite{Oogane06}.
Spin relaxation generates a torque perpendicular to the motion of the spin, resulting in a canting of the precession axis.
Similarly, when a spin current $\jsv$ is applied, the spin relaxation thus was argued to induce a torque  perpendicular to the spin-transfer torque, i.e.,
\begin{align}
 \tauv_{\beta} &\equiv \beta \frac{a^3}{2e} \nv\times(\jsv\cdot\nabla)\nv ,
\end{align}
where $\beta$ is a coefficient representing the effect of spin relaxation \cite{Zhang04,Thiaville05}.

Those effects of non adiabaticity and spin relaxation can be calculated from a microscopic viewpoint \cite{KTS06,TKSLL07}.
Let us go back to the LLG equation for localized spin interacting with conduction electron spin via the $sd$ exchange interaction. 
The total Hamiltonian is $H_{S}-{M}\sum_{\rv} \nv(\rv)\cdot\sigmav-H_{\rm e}$, where $H_S$ and $H_{\rm e}$ are  the Hamiltonian for localized spin and conduction electron, respectively.
The equation of motion for localized spin is given by
\begin{align}
  \dot{\nv} &= \gamma \Bv_{S}\times \nv + \gamma \Bv_{\rm e}\times \nv, \label{LLGfromH}
\end{align}
where $\gamma \Bv_{S} \equiv \frac{1}{\hbar} \frac{\delta H_{S}}{\delta \nv}$ and 
 \begin{align}
\gamma \Bv_{\rm e} & \equiv \frac{1}{\hbar} \frac{\delta H_{\rm e}}{\delta \nv}
 =-\frac{M}{\hbar} \average{\sigmav},
\end{align}
 are the effective magnetic field arising from the localized spin and conduction electron, respectively.
 The field $ \Bv_{\rm e}$ is represented by the expectation value of electron spin density, $\average{\sigmav}$, and all the effects from the conduction electron is included in this field; Equation (\ref{LLGfromH}) is exact if $\average{\sigmav}$ is evaluated exactly.
Field theoretic approach is suitable for a systematic evaluation of the electron spin density. 
We move to a rotated frame where the electron spin is described choosing the local $z$ axis along the localized spin.
In the case we are interested, namely, when the effect of non adiabaticity and damping are weak, 
these effects are treated perturbatively.

The spin density in the laboratory frame is written in terms of the spin in the rotated frame $\tilde{\sev}$ as $\se_{i}=R_{ij}\tilde{\se}_j$, where 
\begin{align}
R_{ij}\equiv 2m_im_j-\delta_{ij},          \label{Rij}                                                                                                                                              \end{align}
 is a rotation matrix, $\mv\equiv (\sin\frac{\theta}{2}\cos\phi, \sin\frac{\theta}{2}\sin\phi, \cos\frac{\theta}{2})$ being the vector which define the unitary rotation.
The perpendicular components (denoted by $\perp$) of electron spin density in the rotated frame 
are calculated as 
\cite{TKS_PR08}
\begin{align}
 \tilde{\se}^\perp &= -\frac{2\rhos}{M} {\cal A}_{{\rm s},0}^\perp -\frac{a^3}{eM}\jsv\cdot{\cal A}_{\rm s}^\perp
  -\frac{\alpha_{\rm sr}}{M} (\zvhat \times {\cal A}_{{\rm s},0}^\perp )
 -\frac{\beta_{\rm sr}}{eM}(\zvhat \times (\jsv\cdot{\cal A}_{\rm s}^\perp)).
 \label{serotresult}
\end{align}
The effect of spin relaxation is included in $\alpha_{\rm sr}$ and  $\beta_{\rm sr}=\hbar/(2M\tau_{\rm s})$, both proportional to the spin relaxation time, $\tau_{\rm s}$ \cite{KTS06}.
The first term of Eq. (\ref{serotresult}) represents the renormalization of the localized spin as a result of electron spin polarization and the second term, induced in the presence of applied spin current, describes the adiabatic spin-transfer torque.
Using the identity 
\begin{align}
R_{ij}({\cal A}_{{\rm s},\mu})_j^\perp&=-\frac{1}{2}(\nv\times\partial_\mu\nv)_i,
& R_{ij}(\zvhat\times {\cal A}_{{\rm s},\mu}^\perp)_j =\frac{1}{2}\partial_\mu\nv_i,                  
\end{align}
 we see that Eq. (\ref{serotresult})  leads to 
\begin{align}
  (1+\rhos a^3)\dot{\nv} &= \alpha\nv\times\dot{\nv} -\frac{a^3}{2e}(\jsv\cdot\nabla)\nv 
  -\frac{\beta a^3}{2e}[\nv\times(\jsv\cdot\nabla)\nv] 
  +\gamma \Bv_{S}\times \nv , \label{LLGfromH2}
\end{align} 
which is the LLG equation taking into account the torque due to electrons.
Here $\alpha\equiv \alpha_{\rm sr}$ and $\beta\equiv \beta_{\rm sr}$, neglecting other origins for Gilbert damping and nonadiabatic torque.

\renewcommand{\phiz}{\phi}
\renewcommand{\betaw}{\beta}
\section{Current-driven  domain wall motion}
Let us briefly discuss dynamics of a domain wall based on the LLG equation (\ref{LLGfromH2}) including the current-induced torques. 
We consider an one-dimensional and rigid wall, neglecting deformation.
For a domain wall to be created, the system must have an easy axis magnetic anisotropy energy. 
We also include the hard-axis anisotropy energy, which turns out to govern the domain wall motion.
Choosing the easy and the hard axises along the $z$ and the $y$ directions, respectively, the anisotropy energy is represented by the Hamiltonian
\begin{align}
 H_{K} &\equiv \sumr \lt[ -\frac{KS^2}{2}\cos^2\theta +\frac{K_\perp S^2}{2}\sin^2\theta\sin^2\phi \rt],
\end{align}
where $K$ and $K_\perp$ are the easy- and hard-axis anisotropy energies (both are positive). 
We need to take into account of course the exchange coupling, which is essential for ferromagnetism, which in the continuum expression reads 
\begin{align}
 H_J & \equiv \intr \frac{JS^2a^2}{2} (\nabla\nv)^2 .
\end{align}
The domain wall solution obtained by minimizing $H_K$ and $H_J$ is Eq. (\ref{DWsolrest}) with $\lambda=\sqrt{J/K}$. 
Considering a rigid wall, we assume that $K \gg K_\perp$.
The low energy dynamics of the wall is then described by two variables (called the collective coordinates),  the center coordinate of the wall, $X(t)$, and the angle $\phi(t)$ out-of the easy plane \cite{Slonczewski72,TKS_PR08}.
The wall profile including the collective coordinates is 
\begin{align}
 n_z(z,t)=\tanh\frac{z-X(t)}{\lambda}, \;\;\;  n_\pm(z,t)\equiv n_x\pm in_y=\frac{e^{\pm i\phi(t)}}{\cosh\frac{z-X(t)}{\lambda}}.
 \label{DWsol}
\end{align}
The equation of motion for domain wall is obtained by putting the wall profile (\ref{DWsol}) in Eq. (\ref{LLGfromH2}) and integrating over spatial coordinate as 
\begin{align}
 \dot{\phi}+\alpha\frac{\dot{X}}{\lambda} =&
 P \frac{\beta}{\lambda}  \jtil \nnr
 \dot{X}-\alpha\lambda \dot{\phi} =& 
 -\vc\sin2\phi + P\jtil ,
 \label{eqs}
\end{align}
where $P\equiv \js/j$ is spin polarization of the current, and both $\vc\equiv \frac{K_{\perp}\lambda S}{2\hbar}$ and $\jtil \equiv    \frac{a^3}{2eS}j$ have dimension of velocity.

When $\beta=0$, the wall velocity when a constant $\jtil$ is applied is easily obtained as \cite{TK04} 
\begin{equation}
\overline{\dot{X}} =
\left\{ \begin{array}{lrr} 
  0  &  \;\;\;\;\; & (\jtil < \jcitil) \\
 \frac{|P|}{1+\alpha^2}\sqrt{\jtil^2-(\jcitil)^2} 
  & \;\;\;\;\; & (\jtil \geq \jcitil )
\end{array}\right.
\end{equation}
and $\jcitil \equiv \frac{\vc}{P}$ is the intrinsic threshold current density \cite{TK04}.
Namely, the wall cannot move if the applied current is lower than the threshold value.
This is because the torque supplied by the current is totally absorbed by the wall by tilting the out of plane angle to be 
$\sin 2\phiz =P \jtil /\vc$ when the current is weak ($|P \jtil /\vc|\leq 1$)  and thus the wall cannot move.  This effect is called the intrinsic pinning effect \cite{TKS_PR08}.
For larger current density, the torque carried by the current induces an oscillation of the angle similar to the Walker's breakdown in an applied magnetic field, and the wall speed also becomes an oscillating function of time. 

\begin{figure}[bt]
\begin{center}
\includegraphics[width=0.5\hsize]{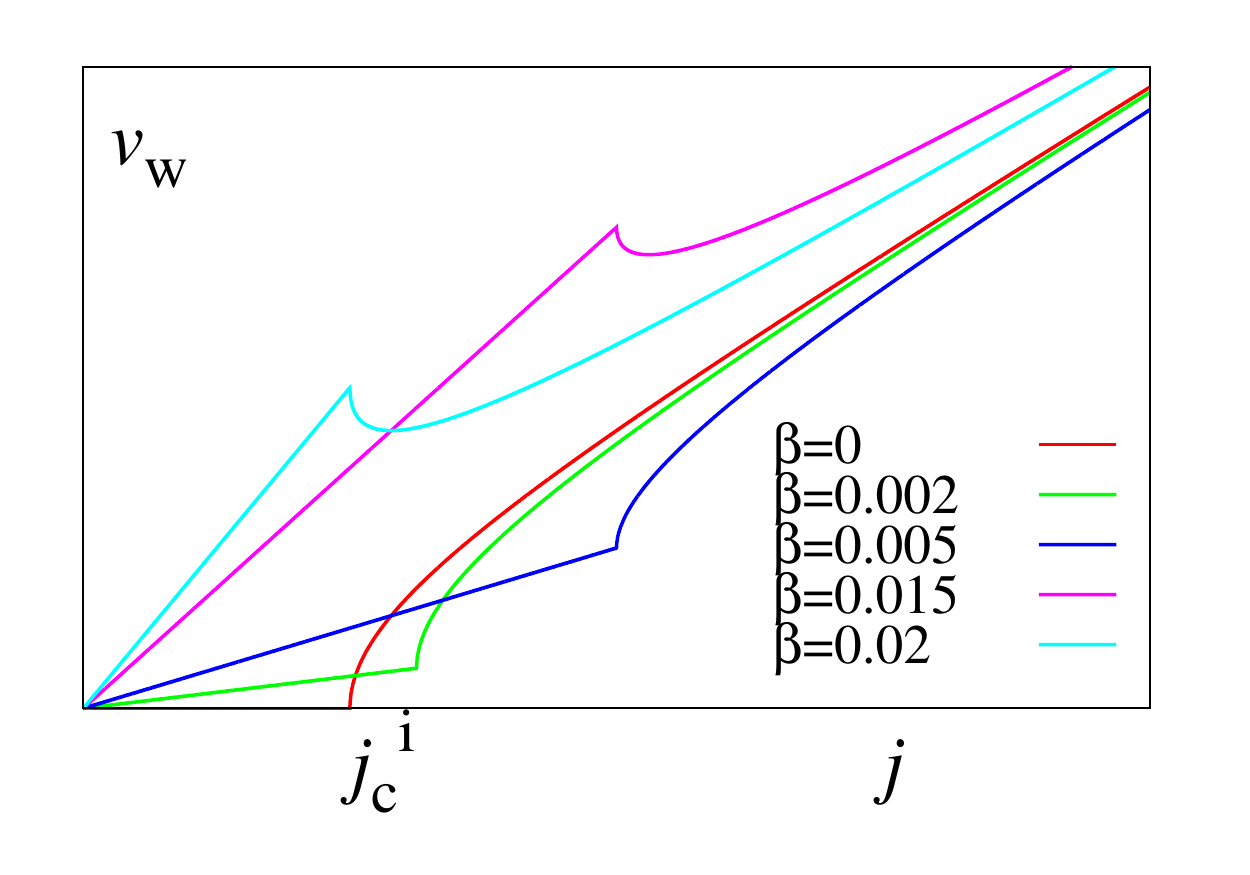}
\caption{ Time averaged wall velocity $v_{w}$ as function of applied spin-polarized current $j$ for $\alpha=0.01$.
Intrinsic pinning threshold $j_{\rm c}^{\rm i}$ exists only for $\betaw=0$. 
The current density where derivative of $v_w$ is discontinuous corresponds to $\jatil$.
\label{FIGvj} }
\end{center}
\end{figure}

When nonadiabaticity parameter $\beta$ is finite, the behavior changes greatly and intrinsic pinning effect is removed and the wall can move with infinitesimal applied current as long as there is no extrinsic pinning. 
In fact, when the applied current density is $ \jtil>\jatil$, where 
\begin{equation}
\jatil \equiv \frac{\vc}{P-\frac{\betaw}{\alpha}},
\label{jatildef}
\end{equation}
the solution of Eq. (\ref{eqs}) is an oscillating function  given by \cite{TKS_PR08} 
\begin{eqnarray}
{\dot{X}}
&=&
\frac{\beta}{\alpha} {\jtil} 
+\frac{\vc}{1+\alpha^2}\frac{\lt(\frac{\jtil}{\jatil}\rt)^2-1}
  {\frac{\jtil}{\jatil}-\sin(2\om t-\vartheta)},
\label{dwvelocity1}
\end{eqnarray}
where 
\begin{align}
\om & \equiv \frac{\vc}{\lambda}\frac{\alpha}{1+\alpha^2}
\sqrt{\lt(\frac{\jtil}{\jatil}\rt)^2-1} , &
\sin \vartheta \equiv \frac{\vc}{(\frac{\betaw}{\alpha}-P)\jtil}.
\end{align} 
The time-average of the wall speed is 
\begin{eqnarray}
{\overline{\dot{X}}}
&=&
\frac{\betaw}{\alpha} \jtil
 +\frac{\vc}{1+\alpha^2}\frac{1}{\jatil}
 \sqrt{\jtil^2-\jatil^2}
.\label{averagewallvelocity}
\end{eqnarray}
For current density satisfying $\jtil<\jatil$, the oscillation in Eq. (\ref{dwvelocity1})
is replaced by an exponential decay in time, and the wall velocity reaches a terminal value of 
 \begin{eqnarray}
\dot{X}
&\ra&
\frac{\beta}{\alpha} {\jtil} .
\label{dwvelocity2}
\end{eqnarray}
The angle of the wall also reaches a terminal value determined by  
\begin{equation}
\sin 2\phiz \rightarrow  
 \left(\frac{\betaw}{\alpha}-P\right) \frac{\jtil}{\vc}. \label{phizterm}
\end{equation}
The averaged wall speed (Eq. (\ref{averagewallvelocity})) is plotted in Fig. \ref{FIGvj}. 

The intrinsic pinning is a unique feature of current-driven domain wall, as the wall cannot move even in the absence of pinning center. 
In the unit of A/m$^2$, the intrinsic pinning threshold is
\begin{align}
 j_{\rm c}^{\rm i}= \frac{eS^2}{Pa^3\hbar}K_\perp \lambda.
\end{align}
For device applications, this threshold needs to be lowered by reducing the hard-axis anisotropy and wall width  \cite{Fukami08}. 
At the same time, the intrinsic pinning is promising for stable device operations. 
In fact,  in the intrinsic pinning regime,  the threshold current and dynamics is insensitive to extrinsic pinning and external magnetic field\cite{TK04}, as was confirmed experimentally \cite{Koyama11}.
This is due to the fact that the wall dynamics in the intrinsic pinning regime is governed by a torque  (right hand side of the second equation of Eq. (\ref{eqs})), which governs the wall velocity $\dot{X}$, while pinning and magnetic field induce force, which governs $\dot{\phi}$; The forces due to sample irregularity therefore does not modify the  motion induced by a torque in the intrinsic pinning regime. 

Experimentally, intrinsic pinning is observed in perpendicularly magnetized materials \cite{Koyama11}, perhaps due to relatively low intrinsic pinning threshold, while materials with in-plane magnetization mostly are in the extrinsic pinning regime governed by the nonadiabatic parameter $\beta$ and extrinsic pinning. 
In this regime, the threshold current of the wall motion is given by \cite{TTKSNF06} 
\begin{align}
 j_{\rm c}^{\rm e} \propto \frac{V_{\rm e}}{\beta},
\end{align}
where $V_{\rm e}$ represents strength of extrinsic pinning potential like those generated by geometrical notches and defects.
Control of nonadiabaticity parameter is therefore expected to be useful for driving domain walls at low current density.

Of recent interest from the viewpoint of low current operation  is to use multilayer structures. 
For instance, heavy metal layers turned out to lower the threshold current by exerting a torque as a result of spin Hall effect \cite{Emori13}, and synthetic antiferromagnets turned out to be suitable for fast domain wall motion at low current \cite{Saarikoski14,YangParkin15}.

\section{Interface spin-orbit effects}
Physics tends to focus on infinite systems or bulk system approximated as infinite, as one of the most important objective of physics is to search for beautiful general law supported by symmetries.
In the condensed matter physics today, studying such 'beautiful' systems seems to be insufficient anymore.
This is because demands to understand physics of interfaces and surfaces has been increasing rapidly as present devices are in  nanoscales  to meet the needs for fast processing of huge data.
Systems with lower symmetry are therefore important subjects of material science today.

Surfaces and interfaces have no inversion symmetry, and this leads to emergence of an antisymmetric exchange interaction (Dzyaloshinskii-Moriya interaction) \cite{Dzyaloshinsky58,Moriya60} in magnetism .
As for electrons, broken inversion symmetry leads to a peculiar spin-orbit interaction, called the Rashba interaction \cite{Rashba60}, whose Hamiltonian is
\begin{align}
{H}_{\rm R} &= i \alphaRv \cdot (\bm{\nabla} \times \bm{\sigma}), \label{RashbaH}
\end{align}
where $\bm{\sigma}$ is the vector of Pauli matrices and $\alphaRv$ is a vector representing the strength and direction of the interaction.
The form of the interaction is the one derived directly from the Dirac equation as a relativistic interaction, but the magnitude can be strongly enhanced in solids having heavy elements compared to the vacuum case.

As is obvious from the form of the Hamiltonian, the Rashba interaction induces electromagnetic cross correlation effects where a magnetization and an electric current are induced by external electric and magnetic field, $\Ev$ and $\Bv$, respectively, like represented as 
\begin{align}
\Mv&=\gamma_{ME}({\bm{{\alpha}}_{\rm R}}\times\Ev), &\jv=\gamma_{jB}({\bm{{\alpha}}}_{\rm R}\times\Bv), \label{spinchargemixing}
\end{align}
where $\gamma_{ME}$ and $\gamma_{jB}$ are coefficients, which generally depend on frequency.
The emergence of spin accumulation  from the applied electric field, mentioned in Ref. \cite{Rashba60}, was studied by Edelstein \cite{Edelstein90} in detail, and the effect is sometimes  called Edelstein effect. 
The generation of electric current by magnetic field or magnetization, called the inverse Edelstein effect \cite{Shen14}, was recently observed in multilayer of Ag, Bi and a ferromagnet \cite{Sanchez13}.

\subsection{Effective magnetic field}
Equation (\ref{RashbaH}) indicates that when a current density $\jv$ is applied, the conduction electron has average momentum of $\pv=\frac{m}{en}\jv$ ($n$ is electron density), and thus an effective magnetic field of 
$
\Bv_{\rm e}=\frac{ma^3}{-e\hbar^2\gamma}\alphaRv\times\jv,
$
acts on the conduction electron spin ($\gamma(=\frac{|e|}{m})$ is  the gyromagnetic ratio).
When the $sd$ exchange interaction between the conduction electron and localized spin is strong, this field multiplied by the the spin polarization, $P$, 
is the field acting on the  localized spin. 
Namely, the localized spin feels a current-induced effective magnetic field of 
\begin{align}
\Bv_{\rm R}=\frac{Pma^3}{-e\hbar^2\gamma}\alphaRv\times\jv . \label{BR}                                                                                                                                               \end{align}

One may argue more rigorously using field theoretic description.
Considering the case of $sd$ exchange interaction stronger than the Rashba interaction, we use a unitary transformation to diagonalize the $sd$ exchange interaction (Eq. (\ref{Ucdef})).
The Rashba interaction in the field representation then becomes 
\begin{align}
{H}_{\rm R} &= -\intr \frac{m}{\hbar e }\epsilon_{ijk} \alpha_{{\rm R},i} R_{kl} \tilde{j}_{{\rm s},j}^l , \label{HRrot}
\end{align}
where $\tilde{j}_{{\rm s},j}^l \equiv -i  \frac{\hbar e }{2m} a^\dagger \nablalr_j \sigma_l a$ is the spin current in the rotated frame, $R_{ij}$ is given in Eq. (\ref{Rij}). 
Terms containing spatial derivatives of magnetization structure is neglected, considering the slowly-varying structure. 
In this adiabatic limit,  spin current is polarized along the $z$ direction, i.e.,  $\tilde{j}_{{\rm s},j}^l =\delta_{l,z}{j}_{\rm s} $. We therefore obtain using $R_{kz}=n_k$, 
\begin{align}
{H}_{\rm R} &= \intr \frac{m}{\hbar e }\jv_{\rm s} \cdot( \alphaRv\times\nv),
\end{align}
which results in the same expression as Eq. (\ref{BR}).

The strength of the Rashba-induced magnetic field is estimated (choosing $a=2$\AA) as  
$B_{\rm R}=2\times 10^{16} \times \alphaR$(Jm)$\js$(A/m$^2$); 
For a strong Rashba interaction $\alphaR=1$ eV\AA\ like at surfaces \cite{Ast07}, $B_{\rm R}=4\times 10^{-2}$ T at $\js=10^{11}$ A/m$^2$.
This field appears not very strong, but is sufficient at modify the magnetization dynamics. 
In fact, for the domain wall motion, when the Rashba-induced magnetic field is along the magnetic easy axis, the field  is equivalent to that of  an effective $\beta$ parameter of
\begin{align}
\beta_{\rm R}=\frac{2m \lambda}{\hbar^2}\alphaR                                        ,
\end{align}
 where $\lambda$ is the wall thickness.
If  $\alphaR=1$ eV\AA, $\beta_{\rm R}$ becomes extremely large like $\beta_{\rm R}\simeq 250$ for $\lambda=50$ nm. Note that $\beta$ arising from spin relaxation is the same order as Gilbert damping constant, namely of the order of $10^{-2}$.
Such a large effective $\beta$ is expected to leads to an extremely fast domain wall motion under current \cite{Obata08,Manchon09}.

Experimentally, it was argued that fast domain wall motion observed in Pt/Co/AlO was due to the Rashba interaction \cite{Miron10}, but the result is later associated with the torque generated by spin Hall effect in Pt layer \cite{Emori13}. 
It was recently shown theoretically that strong Rashba-induced magnetic field works as a strong pinning center when introduced locally, and that this Rashba pinning effect is useful for highly reliable control of domain walls in racetrack memories \cite{TataraDW16}.

\subsection{Rashba-induced spin gauge field}
Since the interaction (\ref{HRrot}) is the one coupling to the spin current, the Rashba interaction is regarded as a gauge field acting on electron spin as far as the linear order concerns.
The gauge field defined by Eq. (\ref{HRrot}) is  
\begin{align}
\Av_{\rm R}\equiv -\frac{m}{e\hbar}(\alphaRv\times\nv).
\label{AvRdef}
\end{align}
Existence of a gauge field naturally leads to an effective electric and magnetic field \cite{Kim12,Nakabayashi14} 
\begin{eqnarray}
  \Ev_{\rm R} = -\dot{\Av}_{\rm R} = \frac{m}{e\hbar}(\alphaRv\times\dot{\nv})  \nnr
  \Bv_{\rm R} = \nablav \times \Av_{\rm R} = - \frac{m}{e\hbar}\nablav\times(\alphaRv\times\nv).
  \label{ERBR}
\end{eqnarray}
In the presence of electron spin relaxation, the electric field has a perpendicular component \cite{Tatara_smf13}
\begin{equation}
  \Ev_{\rm R}' = \frac{m}{e\hbar}\beta_{\rm R}[\alphaRv\times(\nv\times\dot{\nv})] , \label{ERp}
\end{equation}
where $\beta_{\rm R}$ is a coefficient representing the strength of spin relaxation.
For the case of strong Rashba interaction of $\alpha_{\rm R}=3$ eV\AA, as realized in Bi/Ag,  
the magnitude of the electric field is $|E_{\rm R}|=\frac{m}{e\hbar}\alpha_{\rm R}\omega=26$kV/m if the angular frequency $\omega$ of magnetization dynamics is 10 GHz. 
The magnitude of relaxation contribution is $|E_{\rm R}'|\sim260$V/m if $\beta_{\rm R}=0.01$.
The effective magnetic field in the case of spatial length scale of 10 nm is high as well; $B_{\rm R}\sim 260$T. 

\begin{figure}[tb]
  \begin{center}
   \includegraphics[width=0.4\hsize]{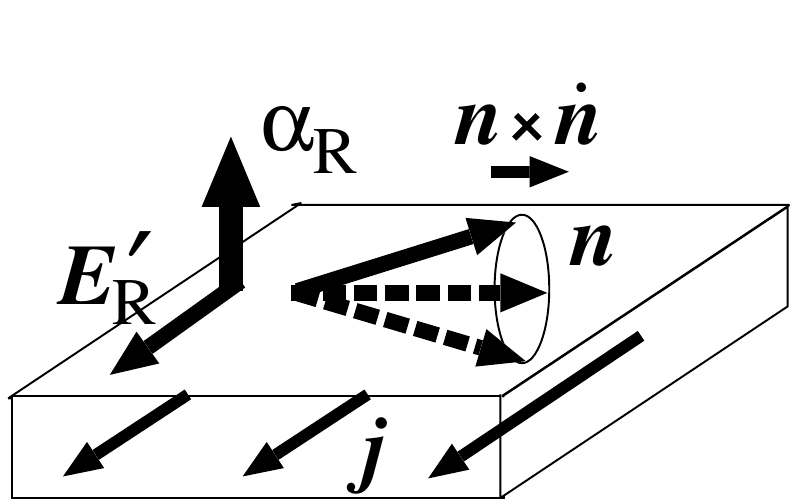}
  \end{center}
\caption{ 
Schematic figure depicting spin relaxation contribution of Rashba-induced spin electric field $\Ev_{\rm R}'$ generated by magnetization precession.
Electric current $\jv$ is induced as a result of motive force $\Ev_{\rm R}'$ in the direction perpendicular to both $\nv\times\dot{\nv}$ and Rashba field $\alphaRv$.
\label{FIGEs_alpha}}
\end{figure}
The Rashba-induced electric fields, $\Ev_{\rm R}$ and $\Ev_{\rm R}'$, are important from the viewpoint of spin-charge conversion.
In fact, results (\ref{ERBR})(\ref{ERp}) indicates that a voltage is generated by a dynamics magnetization if the Rashba interaction is present, even in the case of spatially uniform magnetization, in sharp contrast to the conventional adiabatic effective electric field from the spin Berry's phase of Eq. (\ref{EsBsdef}).
In the case of a think film with Rashba interaction perpendicular to the plane and with a precessing  magnetization,
the component $\Ev_{\rm R}\propto \dot{\nv}$ has no DC component, while 
the relaxation contribution  $\Ev_{\rm R}'$ has a DC component perpendicular to $\overline{\nv\times\dot{\nv}}\parallel \overline{\nv}$.
The geometry of this (spin-polarized) current pumping effect, 
$\jv\propto  \Ev_{\rm R}' \propto  \alphaRv\times \overline{\nv} $,
is therefore the same as the one expected in the case of inverse Edelstein effect (Fig. \ref{FIGEs_alpha}). 
In the present form, there is a difference between the Rashba-induced electric field effect and the system  in Ref. \cite{Sanchez13}, that is, the former  assumes a direct contact between the Rashba interaction and magnetization while   they are separated by a Ag spacer in Ref. \cite{Sanchez13}.
It is expected, however, that the Rashba-induced electric field becomes long-ranged and  survives in the presence of a spacer if we include the electron diffusion processes. 
The  spin-charge conversion observed in junctions will then be interpreted as due to the Rashba-induced electromagnetic field. For this scenario to be justified, it is crucial to confirm the existence of magnetic component,  $ \Bv_{\rm R} $, which can be of the order of 100T.
In the setup of Fig. \ref{FIGEs_alpha},  $ \Bv_{\rm R} $ is along $\overline{\nv}$. 
The field can therefore be detected by measuring ``giant'' in-plane spin Hall effect when a current is injected perpendicular to the plane.

\section{Application of effective vector potential theory}
\subsection{Anomalous optical properties of Rashba conductor}

The idea of effective gauge field  is useful for extending the discussion to include other degrees of freedom, like optical properties.
In fact, the fact that the Rashba interaction coupled with magnetization leads to an effective vector potential $\Av_{\rm R}$ (Eq. (\ref{AvRdef})) for electron spin indicates that the existence of intrinsic spin flow.
Such intrinsic flow affects the optical properties, as  incident electromagnetic waves get Doppler shift when interacting with flowing electrons, resulting in a transmission depending on the direction (directional dichroism), as was theoretically demonstrated in Refs. \cite{Shibata16,Kawaguchi16}.
The magnitude of the directional dichroism for the case of wave vector $\qv$ is given by $\qv\cdot(\alphaRv\times\nv)$. 
The vector $(\alphaRv\times\nv)$ is called in the context of multiferroics the toroidal moment, and it was argued to acts as an effective vector potential for light \cite{Sawada05}. 

\begin{figure}[tb]\centering
\includegraphics[width=0.5\hsize]{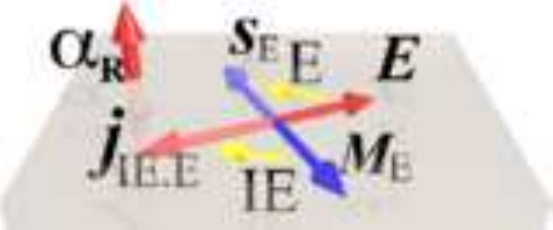}
\caption{ 
Schematic figure showing the cross-correlation effects in the plane perpendicular to the Rashba field $\alphaRv$.
Edelstein effect (E) generates spin density, $\sv_{\rm E}$, from the applied electric field, and inverse Edelstein effect (IE) generates current $\jv_{{\rm IE}\cdot{\rm E}}$ from magnetization $\Mv_{\rm E}$.
\label{FIGEIE}}
\end{figure}

It was shown also that Rashba conductor, even without magnetization, shows peculiar optical properties such as negative refraction as a result of spin-charge mixing effects \cite{Shibata16}.
In fact, spin-charge mixing effects of Eq. (\ref{spinchargemixing})  leads to a current generated by applied electric field, $\Ev$,  given by 
\begin{align}
 \jv_{{\rm IE}\cdot{\rm E}} = -\hbar\gamma\kappa_{\rm EIE} [\alphaRv\times(\alphaRv\times\Ev)], 
\end{align}
where $\kappa_{\rm EIE}$ is a coefficient  
(Fig. \ref{FIGEIE}). 
As it is opposite to the applied field, the mixing effect results in a softening of the plasma frequency as for the $\Ev$ having components perpendicular to $\alphaRv$.
The electric permitivity of the system is therefore anisotropic; Choosing $\alphaRv$ along the $z$ axis, we have 
\begin{align}
\varepsilon_{z}&=1-\frac{\omegaP^2}{\omega(\omega+i\eta)}, 
& 
\varepsilon_{x}=\varepsilon_{y}
=1-\frac{\omegaR^2}{\omega(\omega + i\eta)}, \label{e-perp}
\end{align}
where $\omegaP = \sqrt{{e^2n_{\rm e}}/{\varepsilon_{0}m}}$ 
is the bare plasma frequency ($\nel$ is the electron density), and 
$
  \omegaR \equiv \omegaP\sqrt{1+\Re C(\omegaR)}<\omegaP
$ is the plasma frequency reduced by the spin mixing effect. \cite{Shibata16}
($C(\omega)$ represents the correlation function representing the Rashba-Edelstein effect, and its real part is negative.)
The frequency region $\omegaR<\omega<\omegaP$ is of interest, as 
 the system is insulating ($\varepsilon_z>0$) in the direction of the Rashba field but 
metallic in the perpendicular direction  ($\varepsilon_x<0$).
The dispersion in this case becomes hyperbolic, and the group velocity and phase velocity along $\qv$ can have opposite direction, resulting in negative refraction.
Rashba system is, therefore a natural hyperbolic metamaterial \cite{Narimanov15}.
A great advantage of Rashba conductors are that the metamaterial behavior arises in the infrared or visible light region, which is not easily accessible in fabricated systems. 
For instance, in the case of BiTeI with Rashba splitting of $\alpha=3.85$ eV\AA \cite{Ishizaka11}, 
the plasma frequency is $\omegaP=2.5\times10^{14}$ Hz (corresponding to a wavelength of $7.5\mu$m) for 
$\nel=8\times 10^{25}$ m$^{-3}$ and $\ef=0.2$ eV \cite{Demko12}.
We then have  $\omegaR/\omegaP=0.77$ ($\omegaR=1.9\times 10^{14}$ Hz, corresponding to the wavelength of $9.8\mu$m), and  hyperbolic behavior arises in the infrared regime. 
The directional dichroism arises in the infrared-red light regime \cite{Shibata16}.

\subsection{Dzyaloshinskii-Moriya interaction}
Another interesting effect of spin gauge field pointed out recently is to induce the Dzyaloshinskii-Moriya interaction.
Dzyaloshinskii-Moriya (DM) interaction
is an antisymmetric exchange interaction between magnetic atoms that can arise when inversion symmetry is broken.
In the continuum limit, it is represented as
\begin{align}
H_{\rm DM} \equiv \int d^3r  D^\alpha_i(\nabla_i\bm n\times \bm n)^\alpha,
\label{DMterm}
\end{align}
where $D^a_i$ is the strength, $\alpha$ and $i$ denotes the spin and spatial direction, respectively.
It was recently discussed theoretically that the interaction is a result of Doppler shift due to an intrinsic spin current generated by broken inversion symmetry \cite{Kikuchi16}.
In fact, spin current density, $\js$, which is odd and even under spatial inversion and time-reversal, respectively, is induced by spin-orbit interaction in systems with broken inversion symmetry. 
Spatial variation of localized spins observed by the flowing electron spin is then described by a covariant derivative, 
\begin{align}
 \mathfrak D_i\nv &= \nabla_i\nv+\eta(\jv_{{\rm s},i}\times\nv), \label{covariant}
\end{align}
where $\eta$ is a coefficient. 
This covariant derivative leads to the magnetic energy generated by the electron of 
$(\mathfrak D_i\nv)^2=(\nabla\nv)^2+2\eta \sum_i\jv_{{\rm s},i}\cdot(\nv\times\nabla_i\nv)+O(\eta^2)$.
We see that the second term proportional to $\js$ is the DM interaction, and thus the coefficient is 
$D^a_i \propto j_{{\rm s},i}^\alpha$. 

More rigorous derivation is performed by deriving an effective Hamiltonian.
The electrons interacting strongly with localized spin is described by a Lagrangian (\ref{Lezexpression2}), where 
 ${\cal A}_{{\rm s},i}^\alpha$ is an SU(2) gauge field describing the spatial and temporal variation of localized spin.
To discuss DM interaction, we include a spin-orbit interaction with broken inversion symmetry,
\begin{align}
  H_{\rm so}&= \intr \frac{i}{2}c^\dagger \biggl[\lambda_i\cdot\sigmav \overleftrightarrow{\nabla}_i\biggr] c,
\label{Hso}
\end{align}
where $\lambdav$ is a vector representing the  broken inversion symmetry.
(Multiorbital cases are treated similarly \cite{Kikuchi16}.)
As is obvious from this form linear in spatial derivative and Pauli matrix, the spin-orbit interaction generates a spin current proportional to $\lambdav$. 
From Eq. (\ref{Lezexpression2}), the effective Lagrangian for localized spin to the linear order in derivative is
\begin{align}
\Heff = & \int d^3r  \sum_{i a} \tilde{j}_{{\rm s},i}^a \Acal^a_{{\rm s},i} ,
\label{Heff}
\end{align}
where $\tilde{j}_{{\rm s},i}^a\equiv \average{ \hat{\tilde{j}}_{{\rm s},i}^a }$ is the 
expectation value of the spin current density in the rotated frame.
In terms of the spin current in the laboratory frame, ${j}_{{\rm s},i}^{a}$, the effective Hamiltonian reads 
\begin{align}
\Heff = & \int d^3r  D_i^a (\nabla_i \bm n\times \bm n)^a ,
\label{Heff2}
\end{align}
where 
\begin{align}
        D_i^a \equiv {j}_{{\rm s},i}^{\perp, a},
\label{DMdef}
\end{align}
and ${j}_{{\rm s},i}^{\perp, a}$ is a component of ${j}_{{\rm s},i}^{a}$ perpendicular to the local magnetization direction, $\nv$.
We therefore see that the DM coefficient is indeed given by the 
expectation value of the spin current density of the conduction electrons.  
The first principles calculation based on this spin current expression turns out to have advantage of shorter calculation time than previous methods\cite{Katsnelson10,Freimuth14} by evaluating twist energy of magnetization \cite{Kikuchi16}.

It has been noted that spin wave dispersion is modified in the presence of DM interaction, resulting in Doppler shift of spin waves \cite{Iguchi15,Seki16}. 
The spin wave Doppler shift is natural from our physical interpretation of DM interaction, as DM interaction itself is a consequence of flowing electron spin current.

\section{Summary }

We have discussed various magnetic and electron transport properties in metallic ferromagnets from the view points of effective gauge field.
The concept of gauge field turned out to be highly useful to describe novel electromagnetic cross correlation effects  and optical properties.


\end{document}